\documentclass{article}

\usepackage{changepage}

\usepackage[utf8x]{inputenc}

\usepackage{textcomp,marvosym}

\usepackage{fixltx2e}

\usepackage{amsmath,amssymb}

\usepackage{cite}

\usepackage{nameref,hyperref}

\usepackage{caption}

\makeatletter
\renewcommand{\@biblabel}[1]{\quad#1.}
\makeatother

\date{}

\usepackage{lastpage,fancyhdr,graphicx}
\usepackage{epstopdf}

\usepackage{xargs} 
\usepackage{color}

\begin{document}
\vspace*{0.2in}

\begin{flushleft}
{\Large
\textbf\newline{Understanding Predictability and Exploration in Human Mobility} 
}
\newline
\\
Andrea Cuttone\textsuperscript{1*},
Sune Lehmann\textsuperscript{1,2},
Marta C. González\textsuperscript{3}
\\
\bigskip
\textbf{1} DTU Compute, Technical University of Denmark, Kgs. Lyngby, Denmark
\\
\textbf{2} The Niels Bohr Institute, University of Copenhagen, Copenhagen, Denmark
\\
\textbf{3} Department of Civil and Environmental Engineering and Engineering Systems, Massachusetts Institute of Technology, Cambridge, 02139 Massachusetts, United States of America
\\
\bigskip

* ancu@dtu.dk

\end{flushleft}

%%%%%%%%%%%%%%%%%%%%%%%%%%%%%%%%%%%%%%%%%%%%%%%%%%%%%%
\section*{Abstract}
%%%%%%%%%%%%%%%%%%%%%%%%%%%%%%%%%%%%%%%%%%%%%%%%%%%%%%

Predictive models for human mobility have important applications in many fields such as traffic control, ubiquitous computing and contextual advertisement.
The predictive performance of models in literature varies quite broadly, from as high as 93\% to as low as under 40\%.
In this work we investigate which factors influence the accuracy of next-place prediction, using a high-precision location dataset of more than 400 users for periods between 3 months and one year.
We show that it is easier to achieve high accuracy when predicting the time-bin location than when predicting the next place.
Moreover we demonstrate how the temporal and spatial resolution of the data can have strong influence on the accuracy of prediction.
Finally we uncover that the exploration of new locations is an important factor in human mobility, and we measure that on average 20-25\% of transitions are to new places, and approx. 70\% of locations are visited only once.
We discuss how these mechanisms are important factors limiting our ability to predict human mobility.

%%%%%%%%%%%%%%%%%%%%%%%%%%%%%%%%%%%%%%%%%%%%%%%%%%%%%%
\section*{Introduction}
%%%%%%%%%%%%%%%%%%%%%%%%%%%%%%%%%%%%%%%%%%%%%%%%%%%%%%

Billions of personal devices, ranging from in-car GPS to mobile phones and fitness bracelets, connect us to the cloud. 
These ubiquitous interconnections of the physical and the digital world are opening up a host of new opportunities for predictive mobility models. 
Each user of these devices produces rich information that can help us to capture their daily mobility routine. 
This core knowledge, when obtained from massive number of individuals, impacts a wide range of areas such as health monitoring~\cite{lane2011bewell}, ubiquitous computing~\cite{quercia2010recommending,aalto2004bluetooth}, disaster response~\cite{lu2012predictability} or smart traffic management~\cite{ccolak2016understanding}.

In the age of ubiquitous computing, recent contributions to mobility modeling have flourished in computer science~\cite{eagle2006reality,liao2007learning,zheng2011computing}, transportation engineering~\cite{arentze2000albatross,balmer2008agent}, geographic information sciences~\cite{goodchild2007citizens,batty2013new}, and complexity sciences~\cite{song2010limits,simini2012universal,schneider2013daily}. 
While these findings have enhanced our level of understanding of mobility modeling we need further work to tackle the problem of individual predictability. 

Human mobility has been studied using a multitude of proxies (for example Call Detail Records (CDR), GPS, WiFi, travel surveys), and a variety of techniques have been suggested for predictive models, such as Markov chains, Naive Bayes, artificial neural networks, time series analysis.
Studies report varying results for the predictive power of these models, with accuracy as high as 93\% and as low as under 40\%. 
In this paper we set out to uncover the reasons behind these differences in performance by a thorough investigation of the factors that may influence an estimation of mobility predictability.
The key contributions of this paper are:
\begin{enumerate}
  \item We describe the factors that have influenced the various ranges when estimating predictability. These include: (a) Does the analysis concern the upper limit of predictability, or actual next-place prediction? (b) What is the specific formulation of the prediction problem? E.g. is the goal to predict the next location, or is the goal to identify location in the next time-bin? (c) What is the spatial resolution? E.g. is the analysis based on GPS vs. CDR data? (d) What is the temporal resolution e.g. minutes, hours?
  \item We quantify the amount of explorations and locations visited only once, and show that these are key limiting factors in the accuracy of predictions for individual mobility.
  \item We measure the predictive power of a number of contextual features (e.g. social proximity, time, call/SMS).
  \item We study the problem of predictability of human mobility using a novel, longitudinal, high-precision location dataset for more than 400 users.
\end{enumerate}

The rest of the paper is organized as follows.
We first provide an overview of related work in the field of human mobility prediction.
Next, we introduce the dataset and describe the preprocessing steps.
In the subsequent section we describe the baseline models, and compare their performances.
Finally we introduce the exploration prediction problem and report the performance of the predictive models.

%%%%%%%%%%%%%%%%%%%%%%%%%%%%%%%%%%%%%%%%%%%%%%%%%%%%%%
\section*{Related work}
%%%%%%%%%%%%%%%%%%%%%%%%%%%%%%%%%%%%%%%%%%%%%%%%%%%%%%

In a seminal paper Song \textit{et al}.~\cite{song2010limits} investigate the limits of predictability of human mobility, using Call Detail Records (CDR) as proxy for human movement.
In their analysis, the authors discretize location into a sequence of places, and estimate an upper limit for the predictive performance using Fano's inequality on the temporal entropy of visits.
Their results show that for a majority of users, this upper bound is surprisingly high (93\%).
This framework has been further explored to refine the upper limit.
Specifically, Lin et al.~\cite{lin2012predictability} study the effects of spatial and temporal resolution on the predictability limit, Smith et al.~\cite{smith2014refined} consider the spatial reachability constraints when selecting the next place to visit, and obtain a tighter upper bound of 81-85\%, and Lu et al.~\cite{lu2012predictability} analyze the predictability of the population of Haiti after the earthquake in 2010, and show that the upper limit of predictability remains as high as 85\%.

The work described above focuses on the upper limit of predictability based on estimating the entropy of trajectories. 
When the topic is actual prediction performance, some of the most studied models are Markov chains, where the probability of the next location is assumed to depend only on the current location.
Markov chains have been applied to a variety of data sets. 
Lu et al.~\cite{lu2013approaching} applied Markov chain models to CDR-based locations in Cote D'Ivore, with a prediction goal of estimating the last location of the day at the prefecture (county) level.
Under these conditions the models perform extremely well, reaching an accuracy of over 90\%.
In~\cite{song2006evaluating} the authors apply the Markov models to WiFi traces at Darthmouth campus and find that the best performing model is order 2 and has a median accuracy of about $65-72\%$.
Finally, Bapierre et al.~\cite{bapierre2011variable} applied a variable-order Markov chain to the Reality Mining~\cite{eagle2006reality} and Geolife~\cite{zheng2010geolife} datasets.

Another frequently used category of models is naive Bayes, where the probability of next location is factorized as independent probabilities for a number of context variables.
Gao et al.~\cite{gao2012mobile} applied this approach to the Nokia Data Challenge dataset~\cite{laurila2012mobile} using time and location features, and obtained an accuracy of approximately 50\%.
Do et al.~\cite{do2012contextual} applied the same technique but used a larger number of features including also SMS, calls and  Bluetooth proximity, and obtained an accuracy of approximately 60\%.
In a subsequent paper~\cite{do2015probabilistic} the same authors then explore a kernel density estimation approach for improving performance.

A number of more complex methods have also been explored in the literature, including non-linear time series~\cite{scellato2011nextplace}, Principal Component Analysis~\cite{sadilek2012far}, Gaussian Mixtures~\cite{cho2011friendship} and Dynamic Bayesian Networks~\cite{sadilek2012finding}.

While recent work on predictability has resulted in richer methods and incorporated interesting new features such as social contacts, they have not deeply characterized the intrinsic characteristics of human mobility that form the basis for the limitations in predicting the next visited location. 
In this paper we focus on that aspect, showing that in $53$ weeks, individuals visit on average $200$ unique locations, of which $70\%$ of them are visited only once. Despite most of the trips being among $30\%$ of their recurrent locations; the occurrence of an exploration can be predicted with at best $41\%$ of accuracy. Separating the two types of visited locations and improving the ways to predict an exploration would advance the methods in this area.

%%%%%%%%%%%%%%%%%%%%%%%%%%%%%%%%%%%%%%%%%%%%%%%%%%%%%%
\section*{Materials and Methods}
%%%%%%%%%%%%%%%%%%%%%%%%%%%%%%%%%%%%%%%%%%%%%%%%%%%%%%

%%%%%%%%%%%%%%%%%%%%%%%%%%%%%%%%%%
\subsection*{Data description}
%%%%%%%%%%%%%%%%%%%%%%%%%%%%%%%%%%

In this study we analyze a dataset from the Copenhagen Network Study~\cite{stopczynski2014measuring}.
The project has collected mobile sensing data from smartphones for more than 800 students at the Technical University of Denmark (DTU). 
The data sources include GPS location, Bluetooth, SMS, phone contacts, WiFi, and Facebook friendships.
Data collection was approved by the Danish Data Protection Agency, and informed consent has been obtained for all study all participants.

\begin{figure}[!h]
\centering
\includegraphics[width=1\columnwidth]{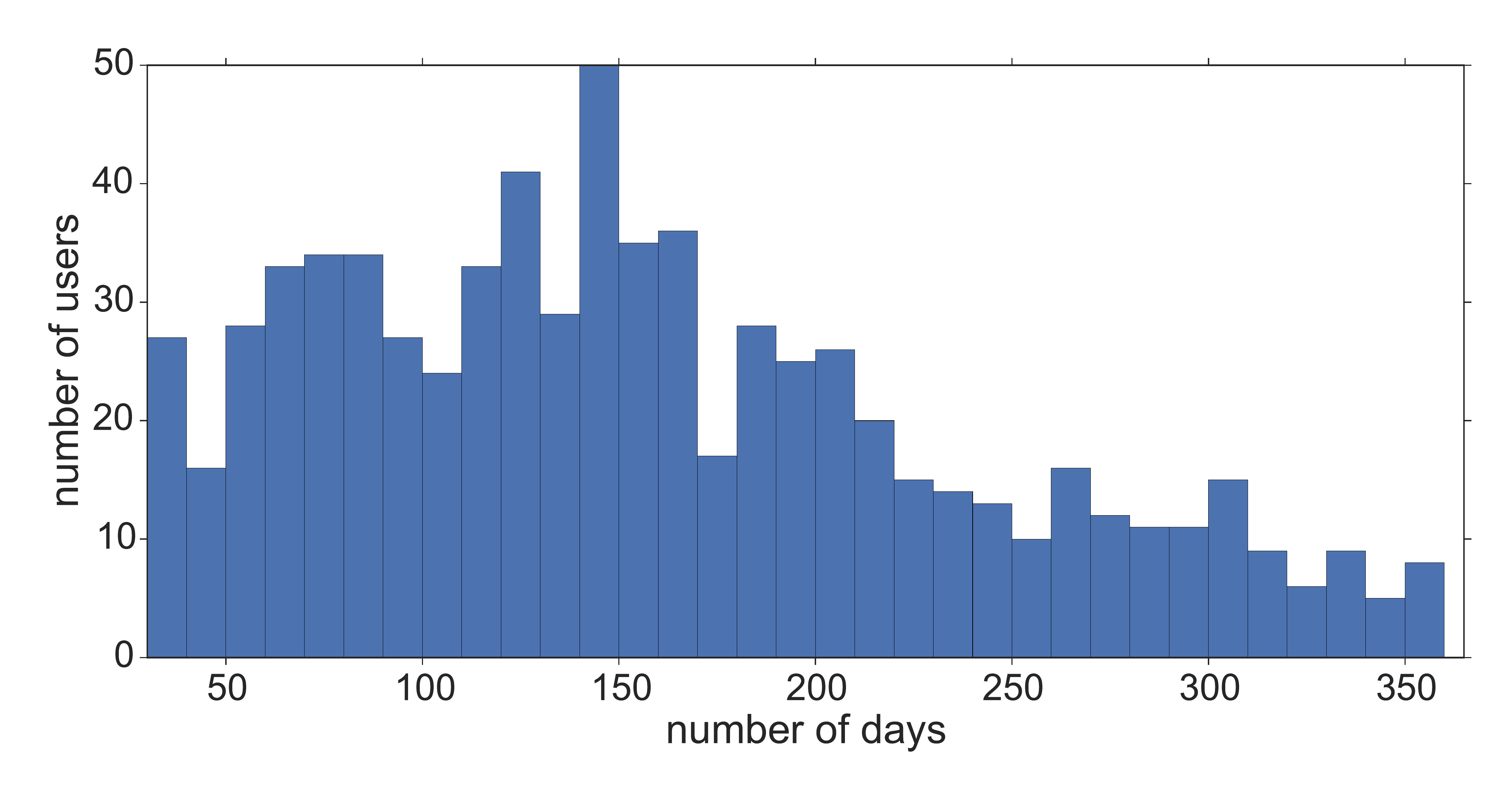}
\caption{
Durations for the periods of collected data for all 454 users.
For each user we select the longest period that has at least one sample in 90\% of the 15-minutes time-bins.
}
\label{fig:periods_lengths}
\end{figure}

For this study we focus on the location data, which is collected by the smartphone with frequency of one sample every 15 minutes.
Each location sample contains a timestamp, a latitude and longitude, and an accuracy value.
The location is determined by the best available provider, either GPS or WiFi, with a median accuracy of $\approx 20$ meters; more than 90\% of the samples are reported to have an accuracy better than $40$ meters.
For individual participants, there may be periods missing data. 
These periods can occur for various reasons, for example due to a drained battery, the phone being switched off, the location probe being disabled, or due to software issues.
Since we are interested in reconstructing mobility histories without large gaps, we select the longest period that has at least one sample in 90\% of the 15-minutes time-bins for each participant.
Moreover we consider only participants that have at least 3 months long period of such contiguous data.
We are left with 454 users, with data collection periods of data ranging from three months to one year. 
Fig.~\ref{fig:periods_lengths} shows the distribution of period lengths.

\begin{figure}[!h]
\centering
\includegraphics[width=1\columnwidth]{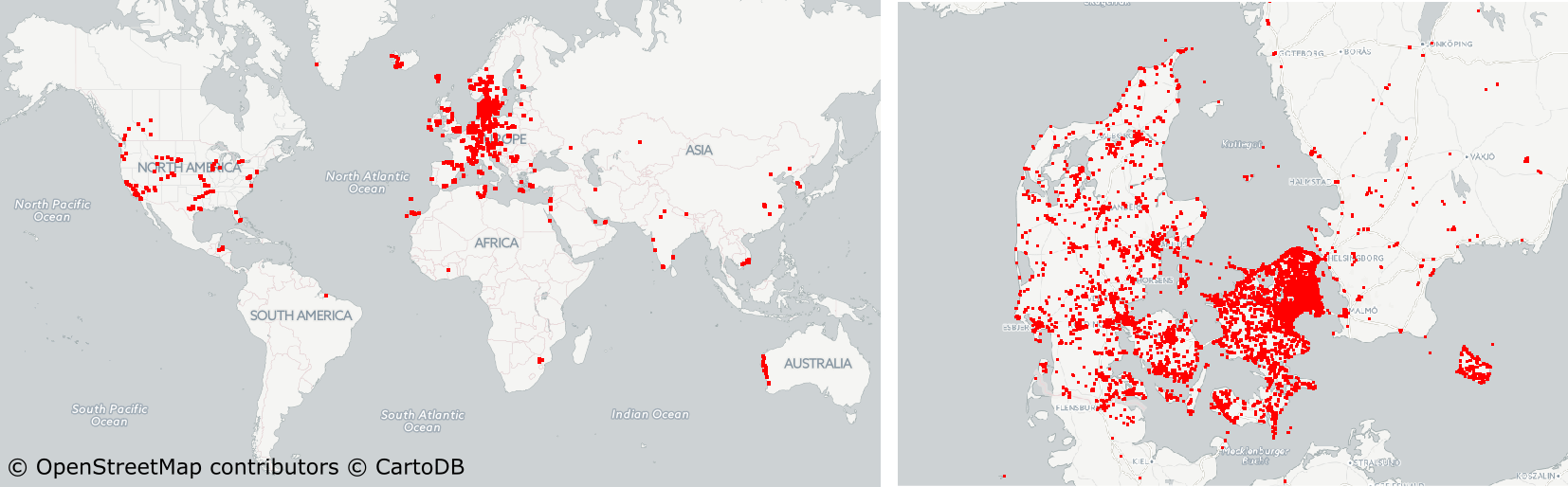}
\caption{
Map of recorded locations across the world (left pane) and in Denmark (right pane).
Each red marker corresponds to a location sample.
}
\label{fig:locmap}
\end{figure}

The data is mainly concentrated in Denmark where the study takes place, but because students use the phones during travel, the dataset spans several other countries as well.
Fig.~\ref{fig:locmap} shows a map of the locations in the world (left pane) and in Denmark (right pane).

In this work we are interested in the location prediction task. 
This task can be broadly stated as follows: given your location history, how well can we predict your future location? 
The specific details of how this question is implemented have a profound impact on the prediction accuracy. 
Below we investigate how various factors, e.g.~spatial and temporal data resolution play a role in determining the reported accuracy for a single underlying dataset.

Because the prediction task can be stated in many different ways, we start the discussion by analyzing different problem formulations. 
In terms of spatial prediction it is possible to discretize space in grid cells, Voronoi cells or define places using a clustering method.
In terms of temporal prediction we could decide to predict a location in the next time-bin, or within a time horizon, or as the next visited place.
In this paper we select two of the most common problem formulations: \textit{next-cell} and \textit{next-place}.
In the \textit{next-cell} formulation we discretize space into grid cells, and we predict the cell in the next time-bin.
In the \textit{next-place} formulation we detect visits to places and we predict the next visited place.
The following sections provide details on the two alternative formulations, and show how each formulation affects the prediction task.

%%%%%%%%%%%%%%%%%%%%%%%%%%%%%%%%%%
\subsection*{next-cell prediction}
%%%%%%%%%%%%%%%%%%%%%%%%%%%%%%%%%%

\begin{figure}[!h]
\centering
\includegraphics[width=1\columnwidth]{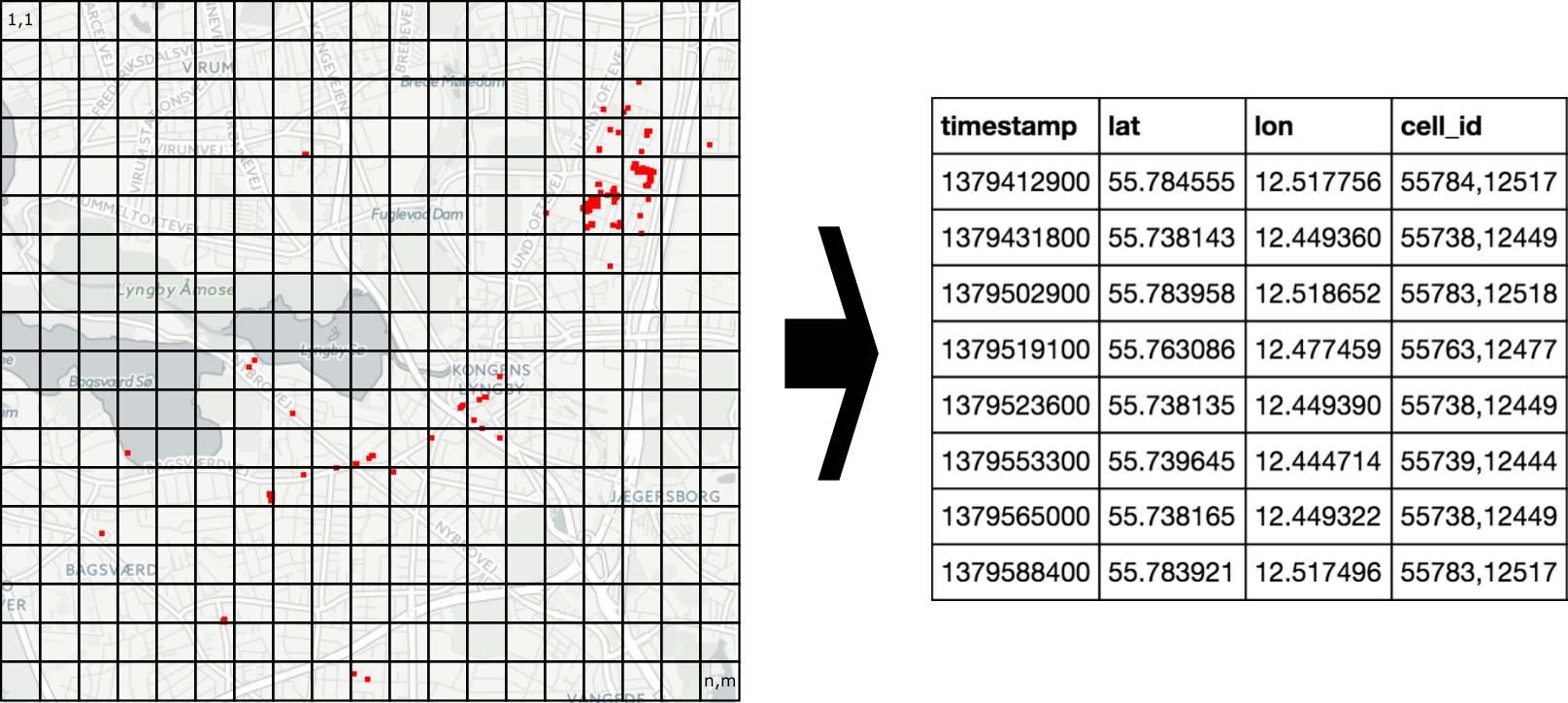}
\caption{Process for converting raw geographical coordinates into sequence of grid cells.
An approximately uniform grid is placed on the map.
For each timestep, the geographical coordinates are converted into the corresponding grid cell ID.
The mobility trace becomes the sequence of visited cell IDs.
}
\label{fig:cell-process}
\end{figure}

In the first problem formulation, we convert geographical coordinates \textit{(lon,lat)} into discrete symbols by placing a uniform grid on the map and retrieving the grid cell id associated with the coordinates.
Specifically, we start by considering a grid of approximate size 50 meters x 50 meters.
At each timestep $\Delta t=15$ minutes, we convert the current \textit{(lon,lat)} into a cell id, therefore producing a sequence of symbols through which we can represent a user's location history.
Fig.~\ref{fig:cell-process} illustrates the process.

\begin{figure}[!h]
\centering
\includegraphics[width=1\columnwidth]{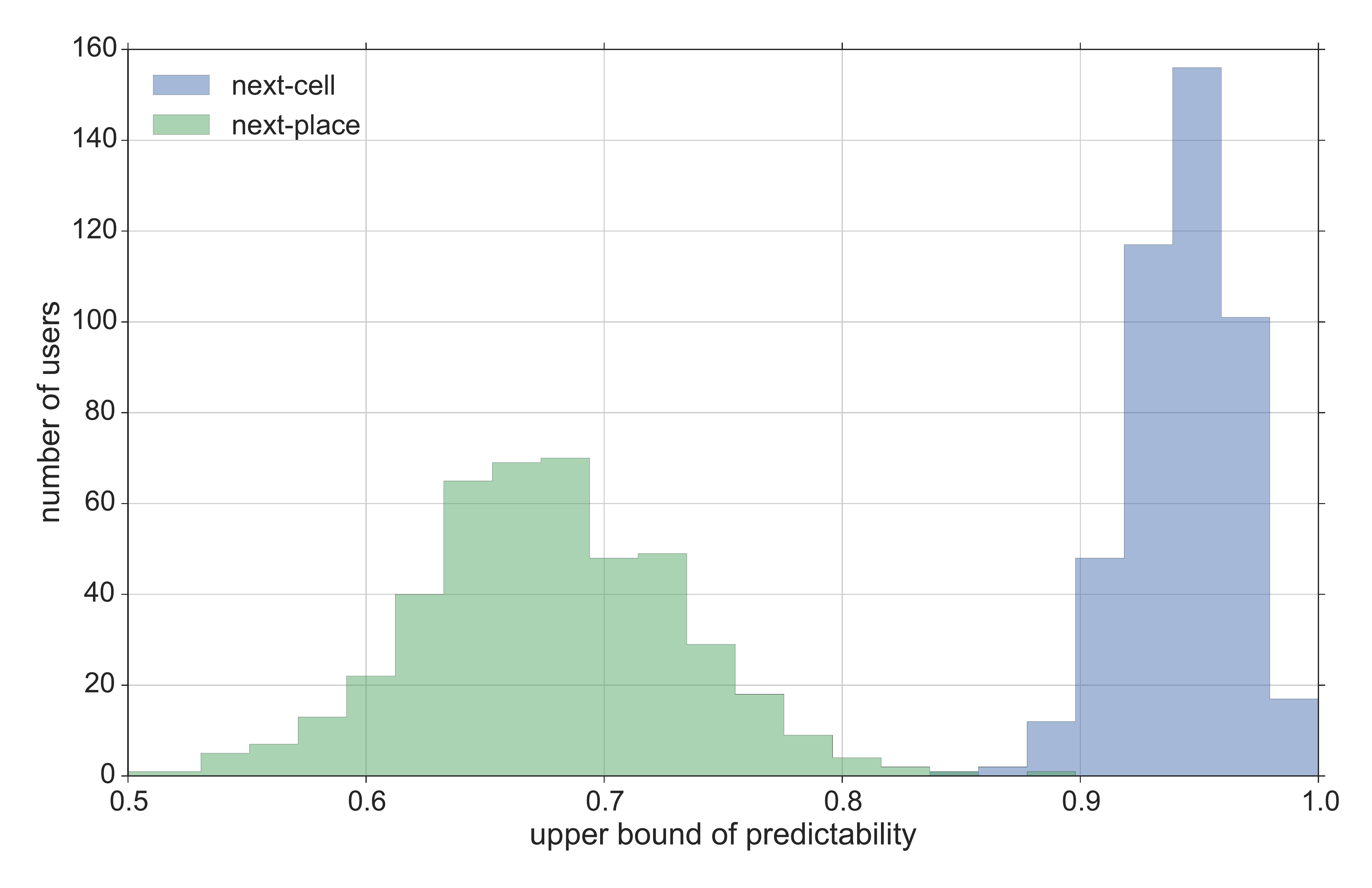}
\caption{Upper bound of predictability for all users for the next-cell and next-place formulations.}
\label{fig:predictability}
\end{figure}

In this formulation, the problem can be restated as follows: given your past cell sequence up to time $t$, which cell will you visit at time $t+\Delta t$?
Before trying to perform any prediction at all, following the process suggested in \cite{song2010limits}, we calculate the theoretical upper limit for the predictability of the cells sequence.
Fig.~\ref{fig:predictability} shows how the maximum predictability for the grid cell formulation is peaked at around 0.95.

We now consider different baseline strategies for next grid cell prediction.
For each of the strategies, we perform prediction in an online manner, by training the algorithm on the data up to timestep $t$, and predicting cell at timestep $t+\Delta t$.
We measure the accuracy as number of correct predictions over the number of total predictions.

\begin{figure}[!h]
\centering
\includegraphics[width=1\columnwidth]{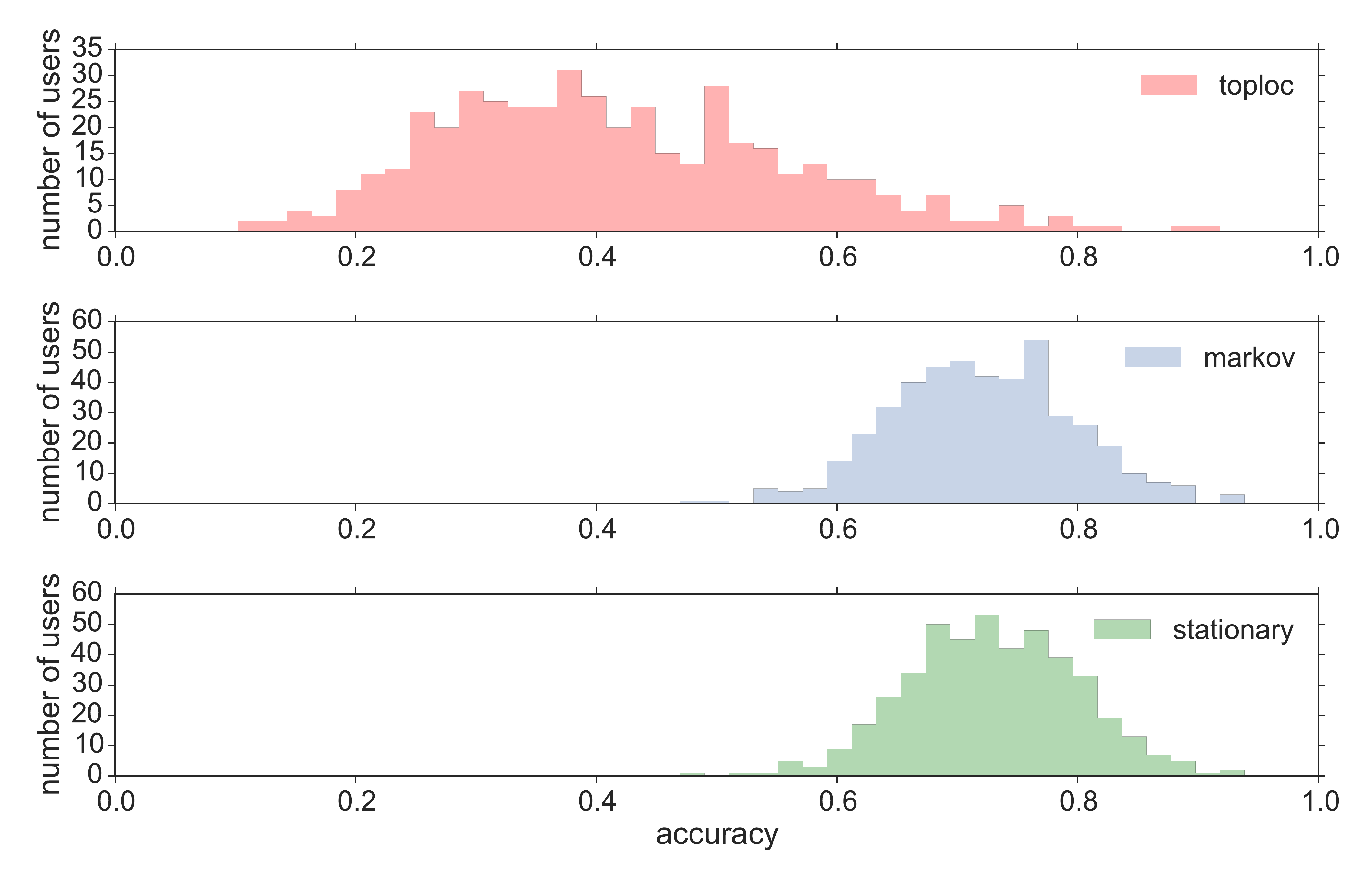}
\caption{Accuracy of the prediction in the next-cell formulation.
The top panel shows the results of the toploc strategy, that is predicting the most common location at each step.
The middle panel shows the accuracy for the Markov chain model.
The bottom panel shows the accuracy for the stationary strategy, that is predicting remaining in the previous cell.
}
\label{fig:timebin-accuracy}
\end{figure}

We first consider the \textit{toploc} strategy, where at each timestep we predict the most frequent symbol in the history so far.
Given the highly stationary nature of most human mobility trajectories, we expect this simple strategy to achieve a relatively high accuracy.
Fig.~\ref{fig:timebin-accuracy} top panel shows the distribution of accuracies for all the users.
The accuracy of the \textit{toploc} is indeed reasonable, peaking at around 0.4. 

We now consider the Markov chain model. 
In this model, the prediction of next state depends only on the current state.
The transition probabilities between locations are estimated based on past transitions in the location history.
For making a prediction, we then consider the transition that has the highest probability among all possible transitions from the current cell.
If the current state has never been seen before, then we have no information about the transition probability to other states.
In this case we fall back and predict the most frequent state.
Again we fit the model in an online manner, updating at each step the transition probabilities and then making a prediction for the next timestep.
Fig.~\ref{fig:timebin-accuracy} middle panel shows the distribution of accuracies for all the users.
The accuracy of the \textit{Markov} model is much higher than toploc, peaking at around 0.7.

\begin{figure}[!h]
\centering
\includegraphics[width=1\columnwidth]{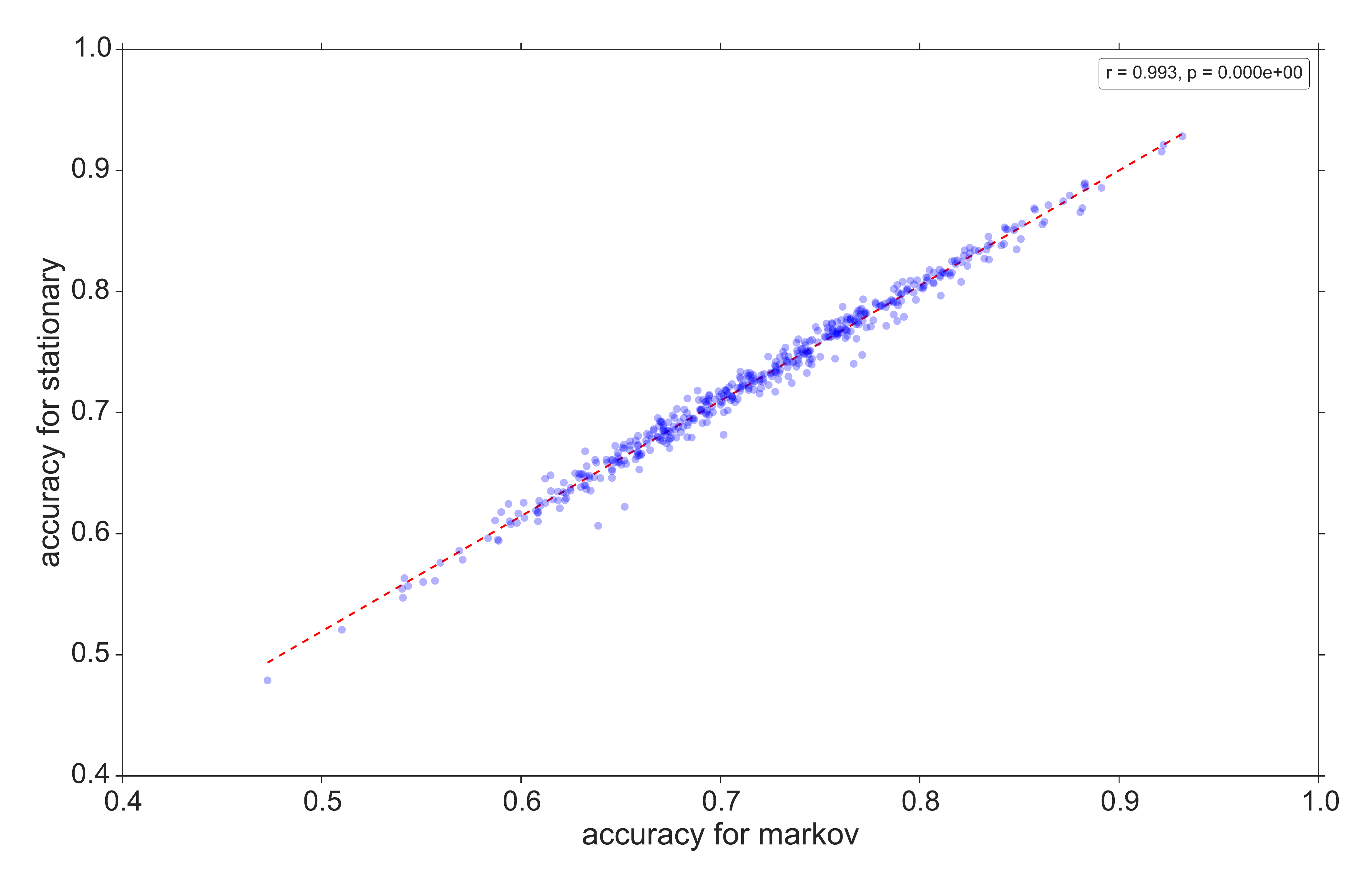}
\caption{Correlation between the accuracy for the Markov model and the stationary model in the next-cell formulation.
}
\label{fig:accuracy-correlation}
\end{figure}

Considering the highly stationary nature of typical trajectories, we hypothesize that a significant part of the Markov prediction power in this formulation comes from self-transitions, that is, the model predicting the user to remain in the same state as in the previous time-bin.
To test this hypothesis, we consider the \textit{stationary} strategy: at each step we predict that the user will remain in the current cell.
Fig.~\ref{fig:timebin-accuracy} bottom panel shows that the distribution of accuracies for \textit{stationary} closely matches the one for \textit{Markov}.
Furthermore Fig.~\ref{fig:accuracy-correlation} shows how the two are very strongly correlated (Pearson's $r = 0.993$, $p < 0.001$).
This strongly suggests that, in this formulation, most of the Markov model power comes from self-transitions, as suspected.

\begin{figure}[!h]
\centering
\includegraphics[width=1\columnwidth]{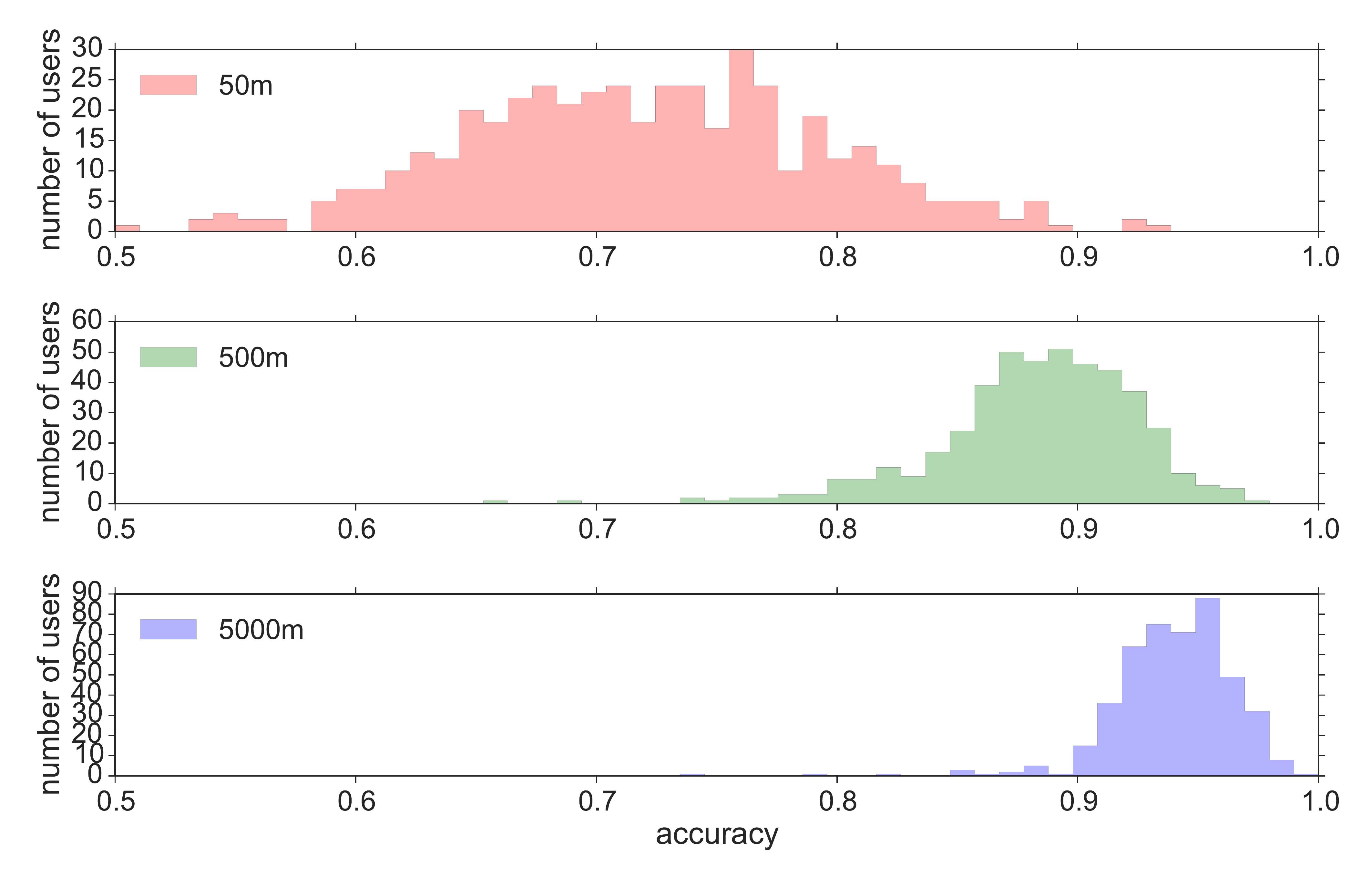}
\caption{
Effect of spatial granularity of the accuracy for the Markov model in the next-cell formulation.
Each panel shows the accuracy for a different spatial bin size: 50m, 500m and 5000m.
Increasing the size of the spatial bins increases the accuracy prediction.
}
\label{fig:cell-accuracy-spatial}
\end{figure}

We now investigate another issue related to this problem formulation.
Intuitively, we expect that the size of our spatial units will influence the accuracy of prediction.
Predicting a user's location with the precision of few meters is intuitively much more difficult that predicting with precision of several kilometers.
In order to examine the effect of spatial resolution, we also consider results for cell size 500 meters and 5000 meters, and apply the \textit{Markov} model.
Fig.~\ref{fig:cell-accuracy-spatial} compares the accuracy for different spatial resolutions.
As expected the accuracy dramatically improves as the spatial size increases.

\begin{figure}[!h]
\centering
\includegraphics[width=1\columnwidth]{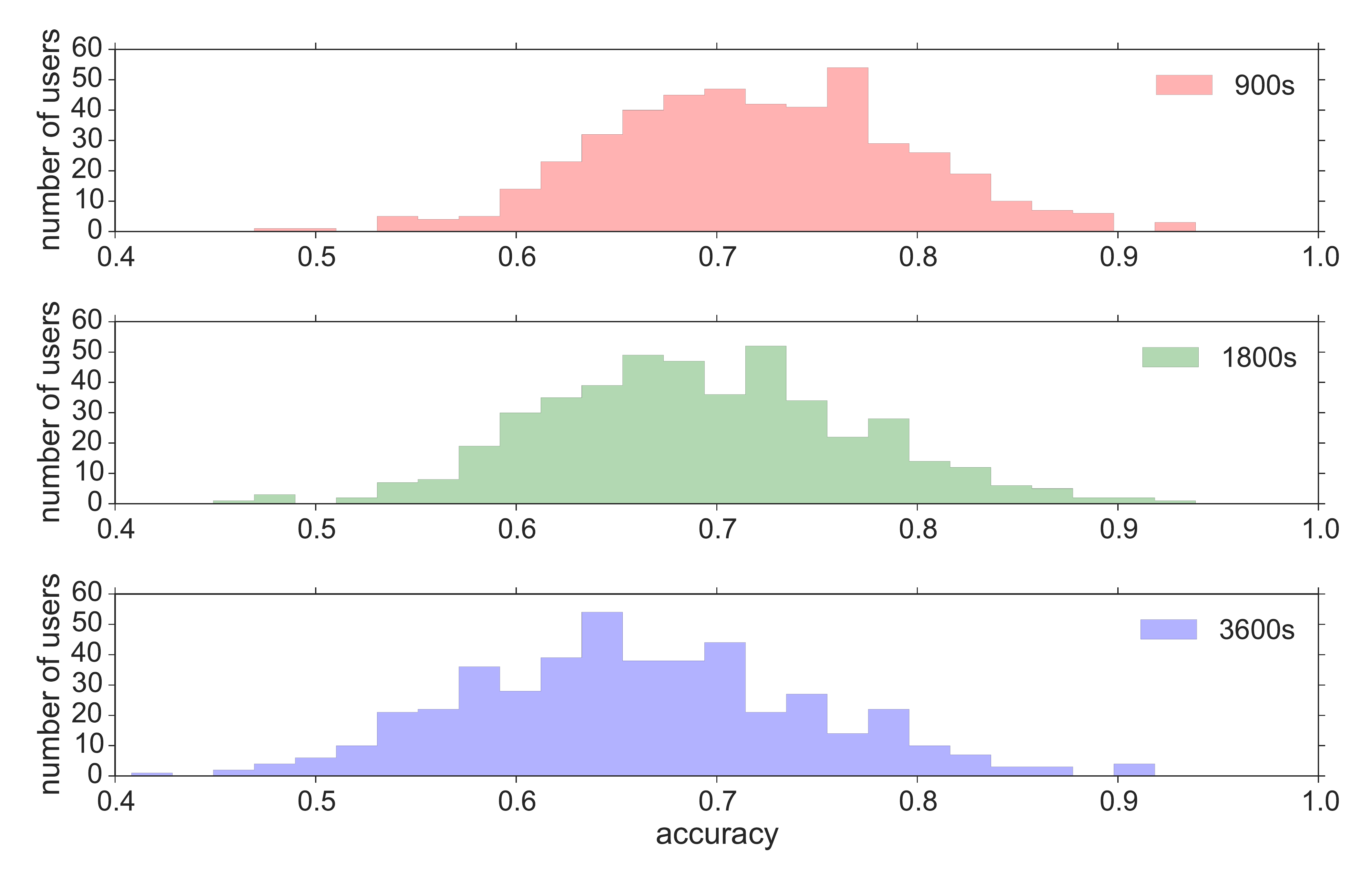}
\caption{
Effect of temporal sampling of the accuracy for the Markov model in the next-cell formulation.
Each panel shows a different temporal bin resolution: 900s, 1800s and 3600s.
Decreasing the temporal resolution in this problem formulation increases the accuracy, since there are more self-transitions.
}
\label{fig:cell-accuracy-temporal}
\end{figure}

Finally we investigate the effect of temporal resolution within this problem formulation.
Our findings above suggest that using a very fine-grained temporal resolution will increase the number of self-transitions, thus driving up the accuracy of the prediction that is mainly able to capture stationarity.
We achieve this by discretizing the location at 50 meters cell size, but varying the temporal time binning to 15 minutes, 30 minutes and 60 minutes, and then running the \textit{Markov} model for each scenario.
Fig.~\ref{fig:cell-accuracy-temporal} compares the accuracy for different temporal resolutions.
As expected, the accuracy is decreased as the time-bins grow larger due fewer self-transitions.

%%%%%%%%%%%%%%%%%%%%%%%%%%%%%%%%%%
\subsection*{next-place prediction}
%%%%%%%%%%%%%%%%%%%%%%%%%%%%%%%%%%

We now consider an alternative problem formulation.
Instead of predicting the cell in the next timestep, we want to predict only when we observe a transition between places, eliminating the possiblity of self-transitions.
In order to do so, we convert the raw GPS locations into a sequence of stops at places.
A large amount of literature has been dedicated to the problem of place detection, such as methods based on WiFi fingerprint~\cite{kang2004extracting}, grid clustering~\cite{zheng2010collaborative}, and kernel density estimation~\cite{thierry2013detecting}.

In this paper we consider the following process, based on density-based clustering approaches such as \cite{zhou2007discovering, zheng2009mining, montoliu2010discovering}.
Each user is treated individually.
We define a ``stop'' as sequence of location-observations where the user has been approximately stationary, that is, the distance between position at time $t$ and $t+\Delta t$ is less than a threshold $\delta = 50$ meters, roughly corresponding to the GPS accuracy.
This produces a sequence of stops, each one with a centroid calculated as the median of the locations coordinates, and a duration equal to the time between the last location and the first location sample.
In order to filter out the short stops during commute, we consider only stops with duration greater than 15 minutes.
The average number of stops per user per day is 2.89 with standard deviation 0.89.

\begin{figure}[!h]
\centering
\includegraphics[width=.7\columnwidth]{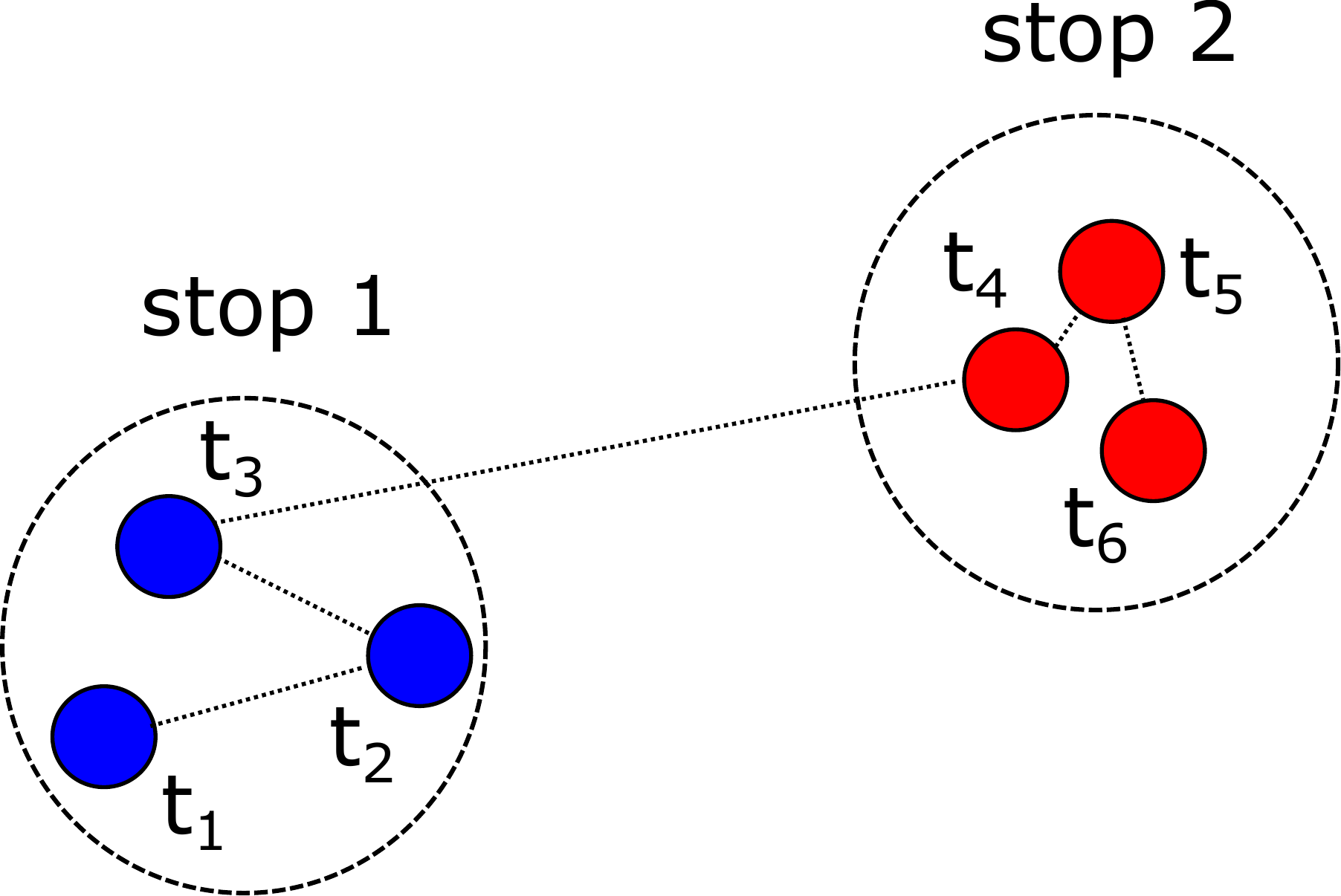}
\caption{Extraction of stops. 
The sequence of location samples $t_1,...,t_6$ are examined sequentially, and are grouped into a stop as long as they are within a distance threshold.
In the example, $t_1, t_2$ and $t_3$ are assigned to the first stop but $t_4$ does not, since it too far away.
Subsequently $t_4$, $t_5$ and $t_6$ are assigned to stop 2.
}
\label{fig:stops-extraction}
\end{figure}

\begin{figure}[!h]
\centering
\includegraphics[width=1\columnwidth]{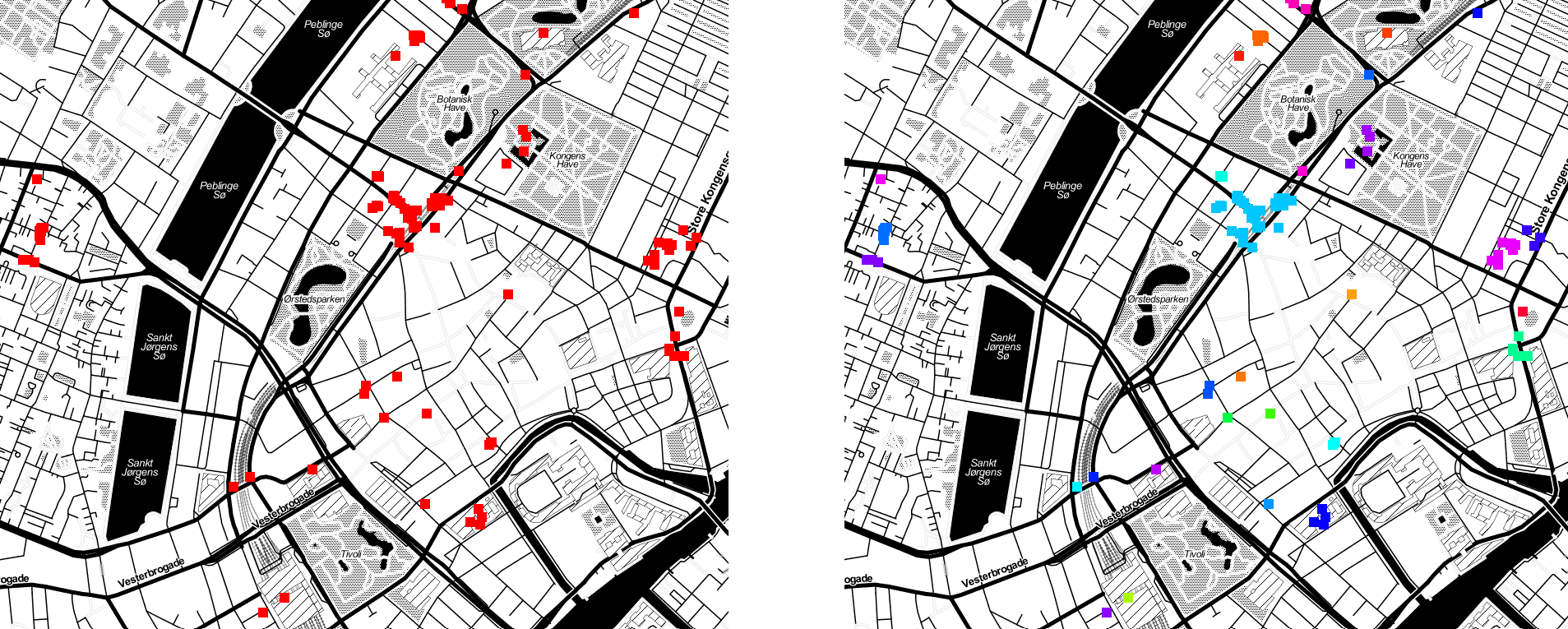}
\caption{Stops are spatially clustered into places using the DBSCAN clustering algorithm, which groups stops according to their haversine distance.
The left panel shows all stops before the algorithm is run, and the right panel shows the assignment of places labels, with each place represented by a different color.}
\label{fig:dbscan}
\end{figure}

We are now interested in grouping stops into places, where a ``place'' is a group of spatially related stops representing a self-contained area such as a building.
In order to do so, we apply the DBSCAN~\cite{ester1996density} clustering to the stops in the geographical coordinate space, using the haversine distance.
We set as parameter the grouping distance $\epsilon = 50$ meters, and \textit{min\_pts} = 2.
This distance threshold is set to produce places of the approximate size of a large building.
The result of the clustering is an assignment of a cluster label to each stop, where the label represents the place that the stop belongs to.
Finally, in order avoid artifacts due to missing samples or noise, we merge multiple consecutive stops at the same place into one.
This process converts the raw location history into a sequence of stops at places.
Fig.~\ref{fig:stops-extraction} and \ref{fig:dbscan} illustrate the complete process of stop detection.

\begin{figure}[!h]
\centering
\includegraphics[width=1\columnwidth]{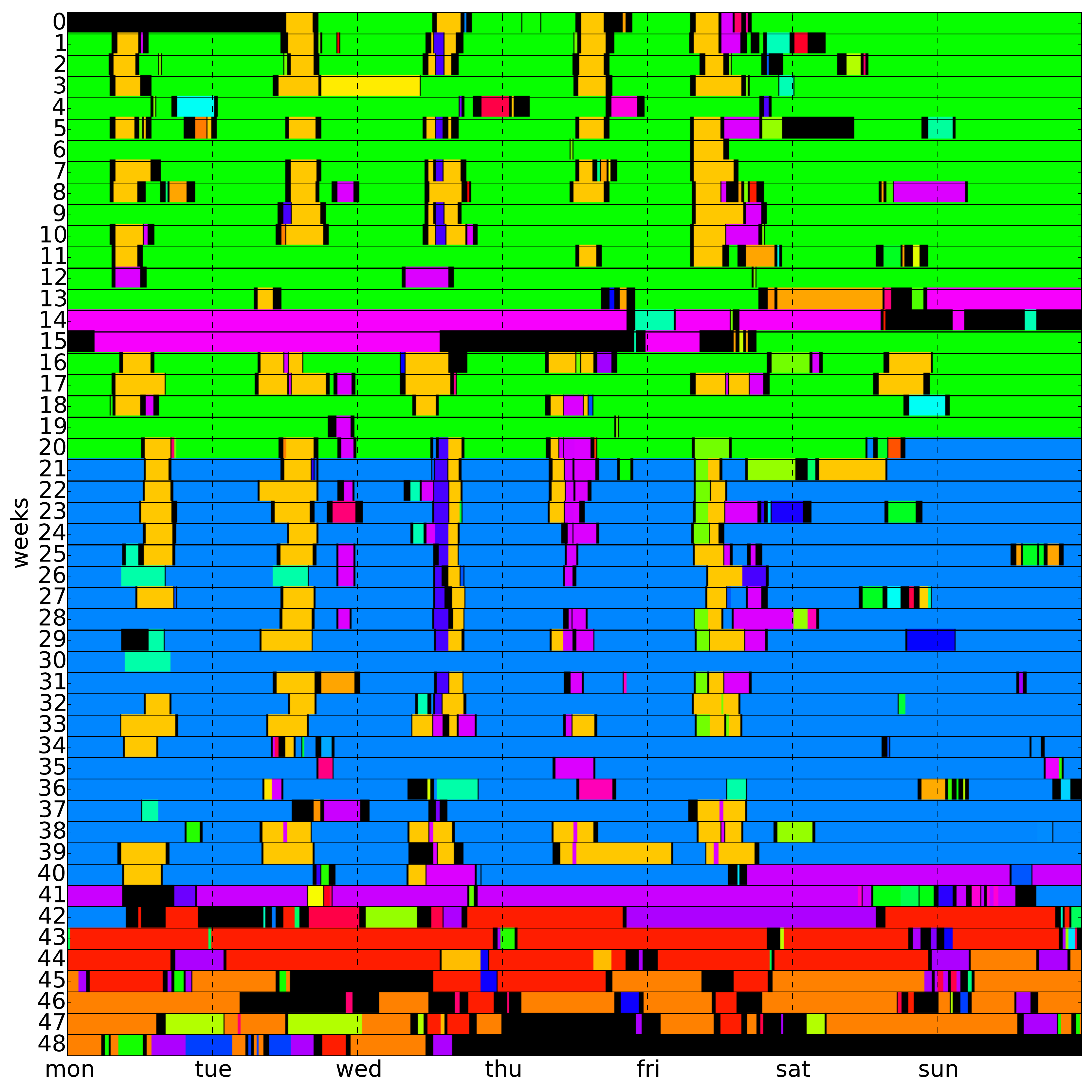}
\caption{Example of the sequence of stops from one user. 
Each row represents a week, from Monday to Sunday.
Each rectangle represents a stop, and its color encodes the corresponding place.
This visualization highlights the complexity of human mobility, with a weekly schedule, periodic returns and irregularities.}
\label{fig:schedule}
\end{figure}

As example result of this process, let us consider the stops and places extracted for a user.
The sequence of stops at places can be represented as a weekly schedule capturing the user's movement patterns.
In Fig.~\ref{fig:schedule} each row represents a week from Monday to Sunday; each place is encoded as a different color.
Inspecting this visualization it immediately possible to spot the periodic patters characterizing human mobility, such as evening returns to the home location, and morning trips to class.
We can also spot many irregularities however, that deviate from the normal schedule: small stops, new explorations, and day-by-day variability.
Finally we can also see a large change in routine starting week 20, where the home location changes.
Each user can be characterized by a similar plot.

The prediction task can now be re-formulated as follows: given a sequence of stops up to step $n$, can we predict your next stop at step $n+1$?
Notice that a key difference from the cell grid formulation is that in this case there are (by definition) no self-transitions; we are interested in the place changes only.

As before, we  start by investigating the upper predictability limit bound.
Fig.~\ref{fig:predictability} shows how the maximum predictability for the stops formulation is peaked at 0.68, significantly lower than what we observe in the grid cells formulation.

\begin{figure}[!h]
\centering
\includegraphics[width=1\columnwidth]{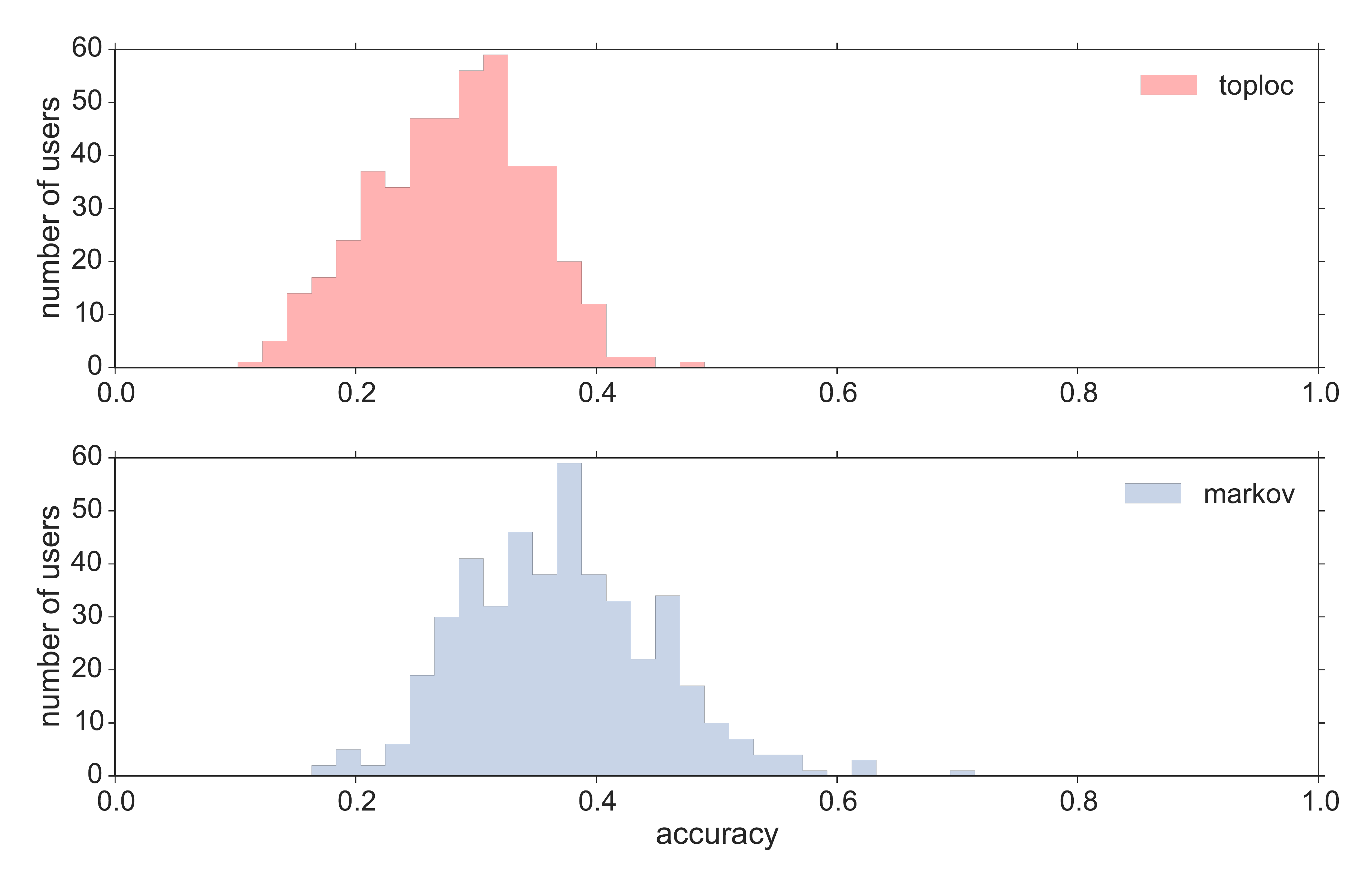}
\caption{
Accuracy for the toploc and Markov models in the next-place formulation.
The accuracy in this formulation is considerably lower than in the next-cell formulation.
}
\label{fig:stops-toploc-Markov}
\end{figure}

We now apply the two prediction strategy \textit{toploc} and \textit{Markov} to this new formulation.
The two models remain conceptually the same, but instead of trying to predict the grid cell at each step, they try to predict the next stop (note that in this formulation we cannot use the \textit{stationary} strategy, as by construction we are interested in transitions to new places).
In this case we also fit each user separately, and we perform the prediction in an online manner.
Fig.~\ref{fig:stops-toploc-Markov} shows the accuracy for both models.
It is evident that the accuracy for these models (around 0.3 for toploc and 0.4 for Markov) is significantly lower in the next-place formulation, indicating that this problem formulation presents a more difficult task.

%%%%%%%%%%%%%%%%%%%%%%%%%%%%%%%%%%
\subsection*{Importance of contextual features}
%%%%%%%%%%%%%%%%%%%%%%%%%%%%%%%%%%

We have investigated how the details of the problem formulation strongly impact the reported accuracy for location prediction tasks. 
We now focus on next-place prediction and study the influence of different contextual features on the prediction task.
We formalize the problem as follows.
At each step, we want to compute the most probable next location given the current location.
We may also want to include other context variables, such as time of the day, day of the week, call activity, or distance from home for example.
In other words we want to compute $P(\hat{L}|c_1,c_2,c_3,...,c_n)$, where $\hat{L}$ is the next location, and $c_1$, $c_2$, $c_n$ are the variables representing different contexts.
For this purpose we use a logistic regression model, and we study the usefulness of various predictor features.
The goal of the model is not to suggest a new state-of-the-art method, but rather to evaluate the importance of individual contextual features.
Specifically, we consider the current location, the time metadata (hour of the day, day of the week, hour of the week, weekend), a `home' binary indication, distance from home, call and SMS activity, and Bluetooth proximity.
Table~\ref{tab:features} provides a summary of the features.

\begin{table}[ht]
    \centering
    \begin{tabular}{|l|p{9cm}|}
    \hline
    \textbf{feature name}  & \textbf{description}                                                                       \\ \hline
    location               & location ID                                                                                \\ \hline
    hour                   & hour of the day (0-23)                                                                     \\ \hline
    weekhour               & hour of the week (0-167)                                                                   \\ \hline
    weekday                & day of the week (0-6)                                                                      \\ \hline
    weekend                & sat/sun (1) or Mon-Fri (0)                                                                 \\ \hline
    explore\_before        & 1 if the previous stop is an exploration, 0 otherwise                                      \\ \hline
    explore\_now           & 1 if the current stop is an exploration, 0 otherwise                                       \\ \hline
    home                   & 1 if the current stop is at the home location (most visited place), 0 otherwise            \\ \hline
    d\_from\_home          & distance from the current stop to the home location                                        \\ \hline
    sms\_received\_30min   & number of SMS received in the 30 min before the current stop timestamp                     \\ \hline
    sms\_sent\_30min       & number of SMS sent in the 30 min before the current stop timestamp                         \\ \hline
    calls\_received\_30min & number of phone calls received in the 30 min before the current stop timestamp             \\ \hline
    calls\_sent\_30min     & number of phone calls sent in the 30 min before the current stop timestamp                 \\ \hline
    bt\_entropy\_30min     & entropy of Bluetooth devices scanned in the 30 min before the current stop timestamp       \\ \hline
    bt\_unique\_30min      & number of unique Bluetooth devices scanned in the 30 min before the current stop timestamp \\ \hline
    \end{tabular}
    \caption{Description of the features used for the logistic regression models.}
    \label{tab:features}
\end{table}

\begin{figure}[!h]
\centering
\includegraphics[width=1\columnwidth]{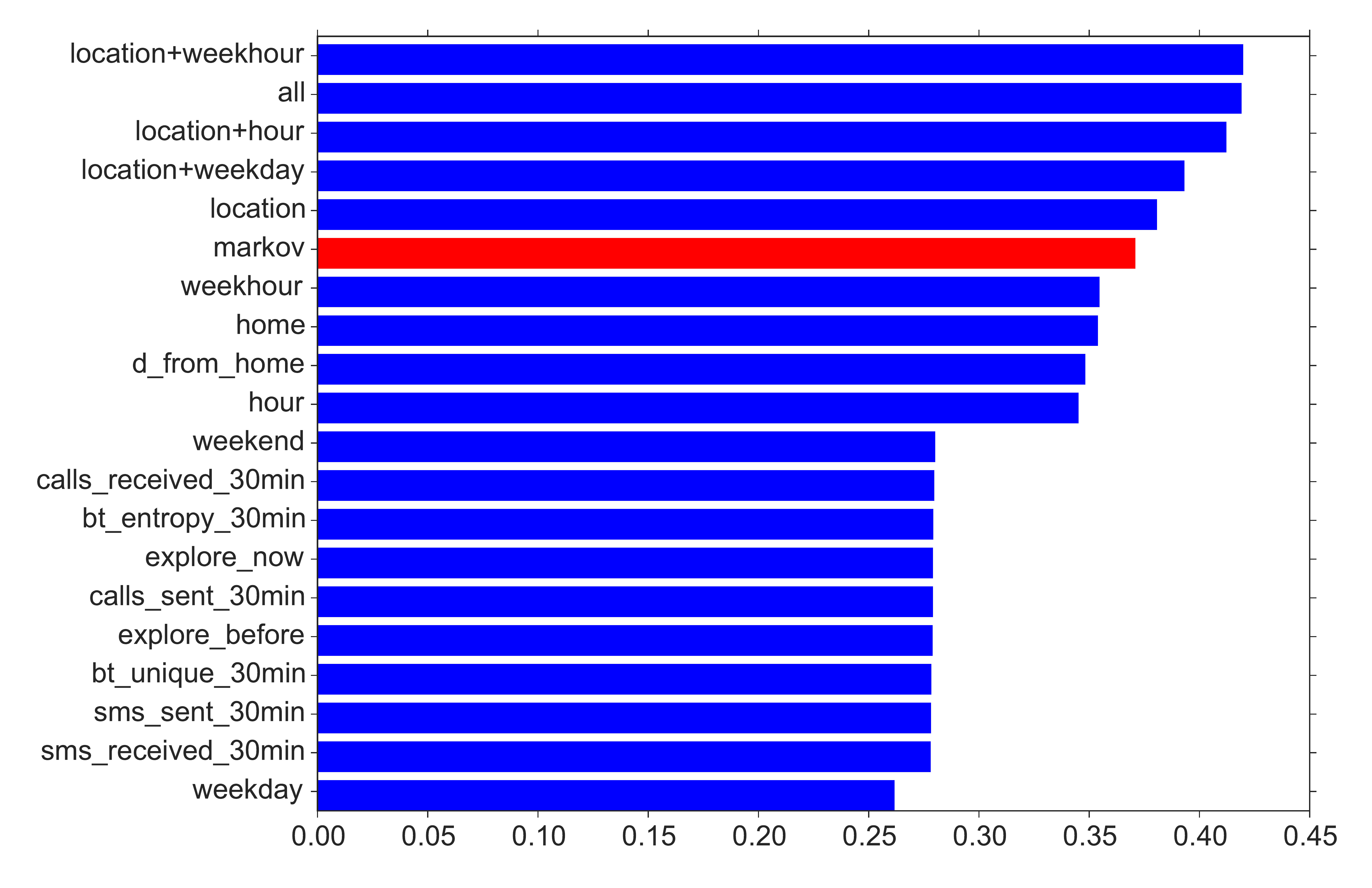}
\caption{
Summary of next-place prediction accuracy for all logistic regression models.
The location and time-related features are the most predictive ones, and outperform the Markov model baseline.
}
\label{fig:stops-acc}
\end{figure}

We model each user separately since we want to perform next-place prediction at the individual level.
As before, we perform an online prediction where we fit the data up to step $n$, and we predict the next location at step $n+1$.
For each user, we fit a logistic regression model using all the individual features separately, and then a model with all features.
Fig.~\ref{fig:stops-acc} shows the accuracy for each of the models, averaged by user.

The Markov chain model baseline is highlighted in red.
Using the current location and time features, the logistic regression model outperforms the Markov chain based model.
Even using the current location only (which is conceptually very similar to a Markov chain model), the logistic regression shows stronger performance, likely due to the explicit optimization of the model.
It is also interesting that other context variable such as call and SMS data have little predictive power in this model formulation.
The most complex model that considers all features is practically identical in performance to the model using only current location and hour of the week.

Although the logistic regression model does improve the accuracy over the Markov model, the absolute value of accuracy is remains low (below 45\%).
We therefore investigate the possible reasons of this difficulty in prediction.

%%%%%%%%%%%%%%%%%%%%%%%%%%%%%%%%%%
\subsection*{Understanding the set of location states}
%%%%%%%%%%%%%%%%%%%%%%%%%%%%%%%%%%

\begin{figure}[!h]
\centering
\includegraphics[width=1\columnwidth]{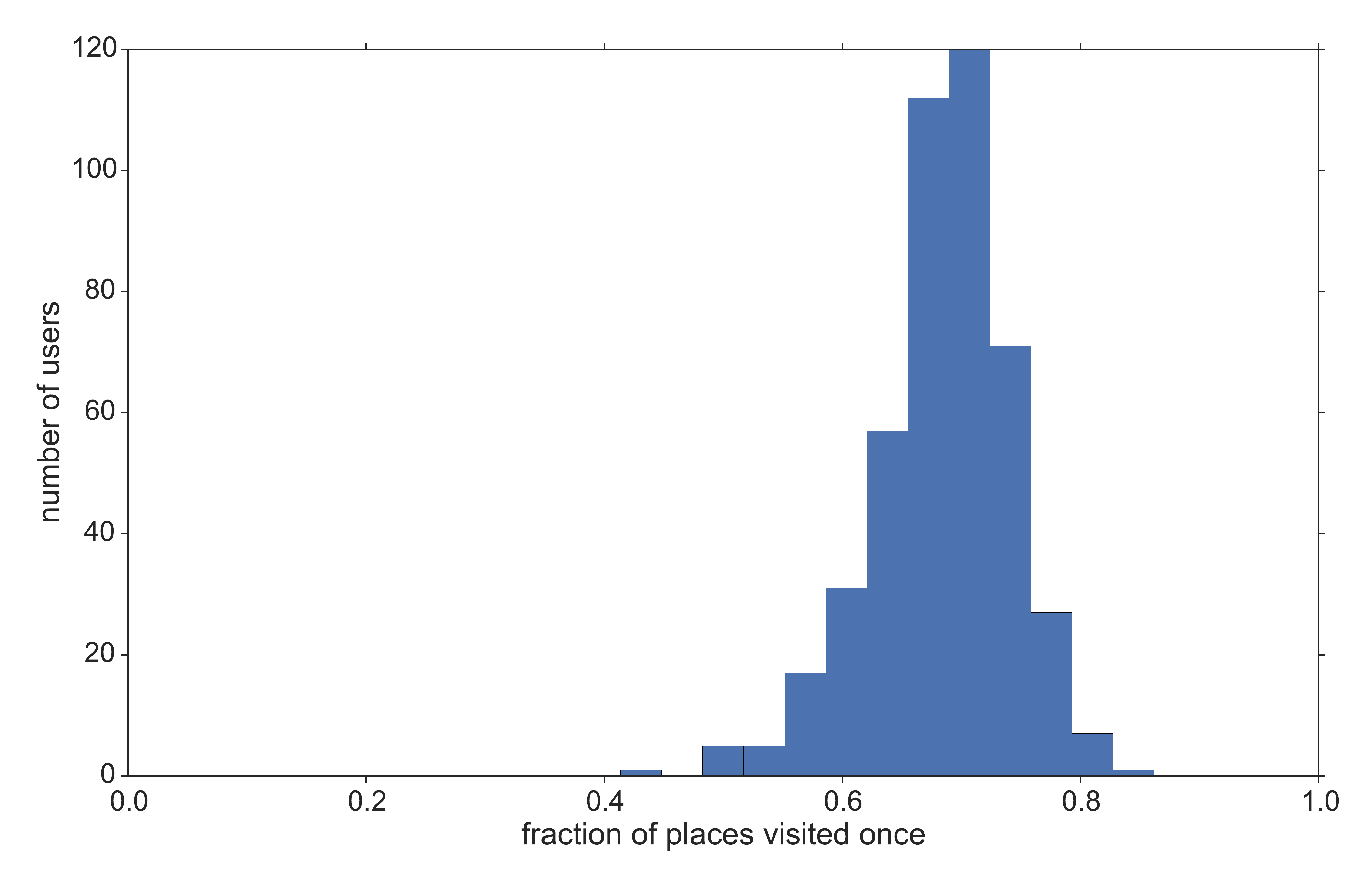}
\caption{
For each user we measure the fraction of places visited only once.
This fraction is surprisingly large, as for each user on average 70\% of the places were visited only once.
}
\label{fig:visited_once}
\end{figure}

It is well known that the majority of individuals tend to spend most of the times at very few places such as home and work, and only sporadically visit other places.
This phenomenon has been described using concepts such as preferential return~\cite{song2010modelling}, heavy-tailed stay times and return rate based on the number of visits~\cite{gonzalez2008understanding}.
For the location prediction tasks, the consequence is that the target classes are very unbalanced, which implies that most records belong to very few classes and most classes are represented by only few records.
To illustrate this issue, we consider the extreme case of places visited only once.
Fig.~\ref{fig:visited_once} shows that surprisingly this fraction is quite large (0.7). 
This fact is, in large part, the central reason behind the difficulty of the prediction task.

As we shall see below, another central challenge is not just that our population visits a large number of different places, but also that many new places are discovered over time.
We consider a stop at a location as ``exploration'' if this place has not been seen in the location history so far for a given user. 
In other words, this place is being visited for the first time by the user.
To express this formally, we consider the sequence of stops $s_1, s_2, s_3, ..., s_n$ for each user .
We consider a stop $s_i$ as \textit{return} $(Y=0)$ if $s_i$ has been seen before in the location history, that is there exists a stop $s_j = s_i$ for $1 \leq j < i$.
Otherwise we consider stop $s_i$ an \textit{exploration} $(Y=1)$, that is the place $s_i$ is visited for the first time at step $i$.
For example given a location sequence A B A C B C, the target variable exploration would have values 1 1 0 1 0 0.

\begin{figure}[!h]
\centering
\includegraphics[width=1\columnwidth]{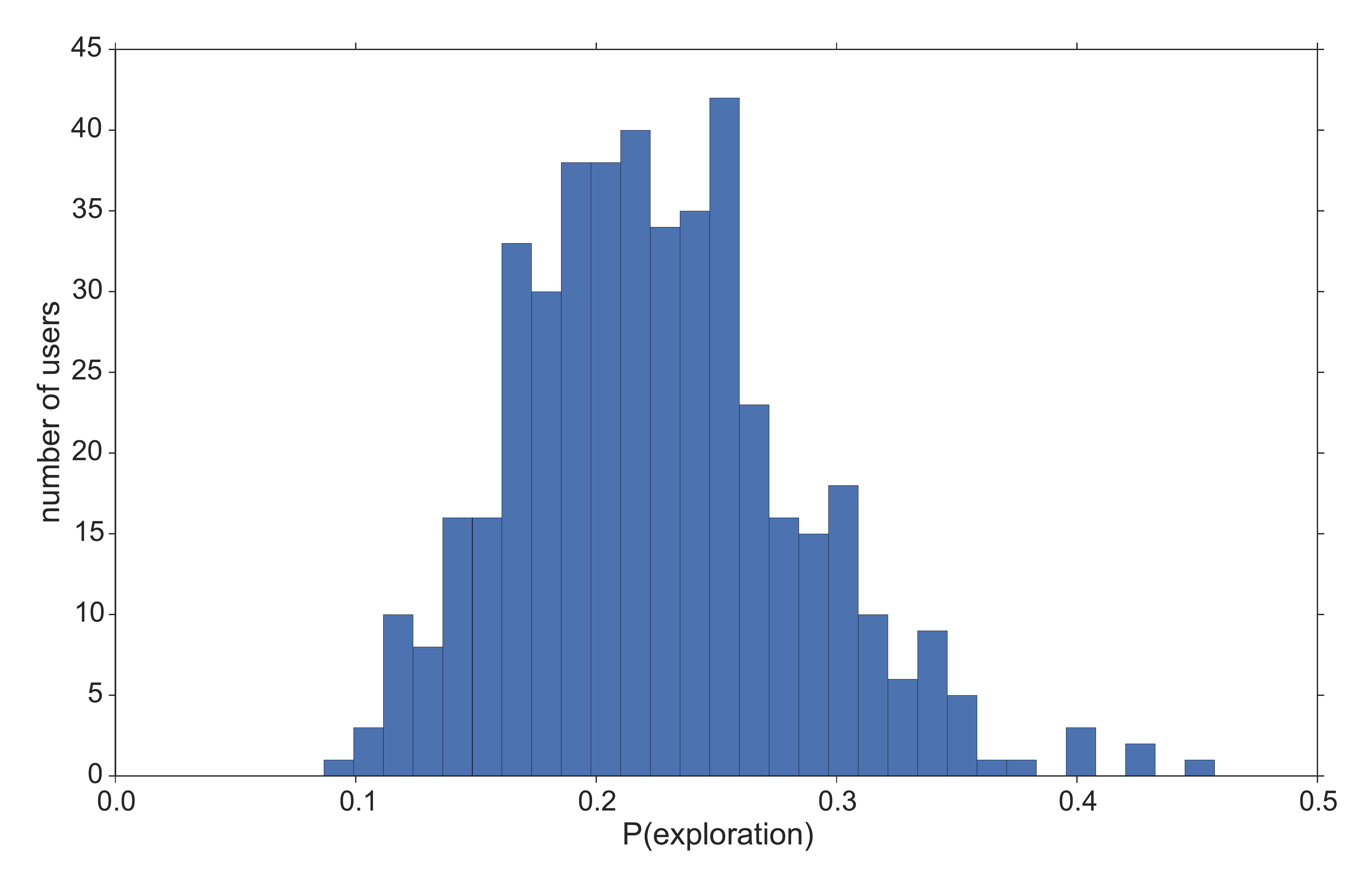}
\caption{
The probability of exploration estimated as fraction of explorations over the number of stops per user.
Surprisingly this probability is quite large, meaning that on average users discover a new place every 4 or 5 stops.
}
\label{fig:p_exploration}
\end{figure}

We can then estimate the probability of exploration as fraction of explorations over the number of stops.
To our surprise, this probability is particularly large: between 0.2 and 0.25 (Fig.~\ref{fig:p_exploration}).
This implies that most users discover a new place every 4 or 5 stops.

\begin{figure}[!h]
\centering
\includegraphics[width=1\columnwidth]{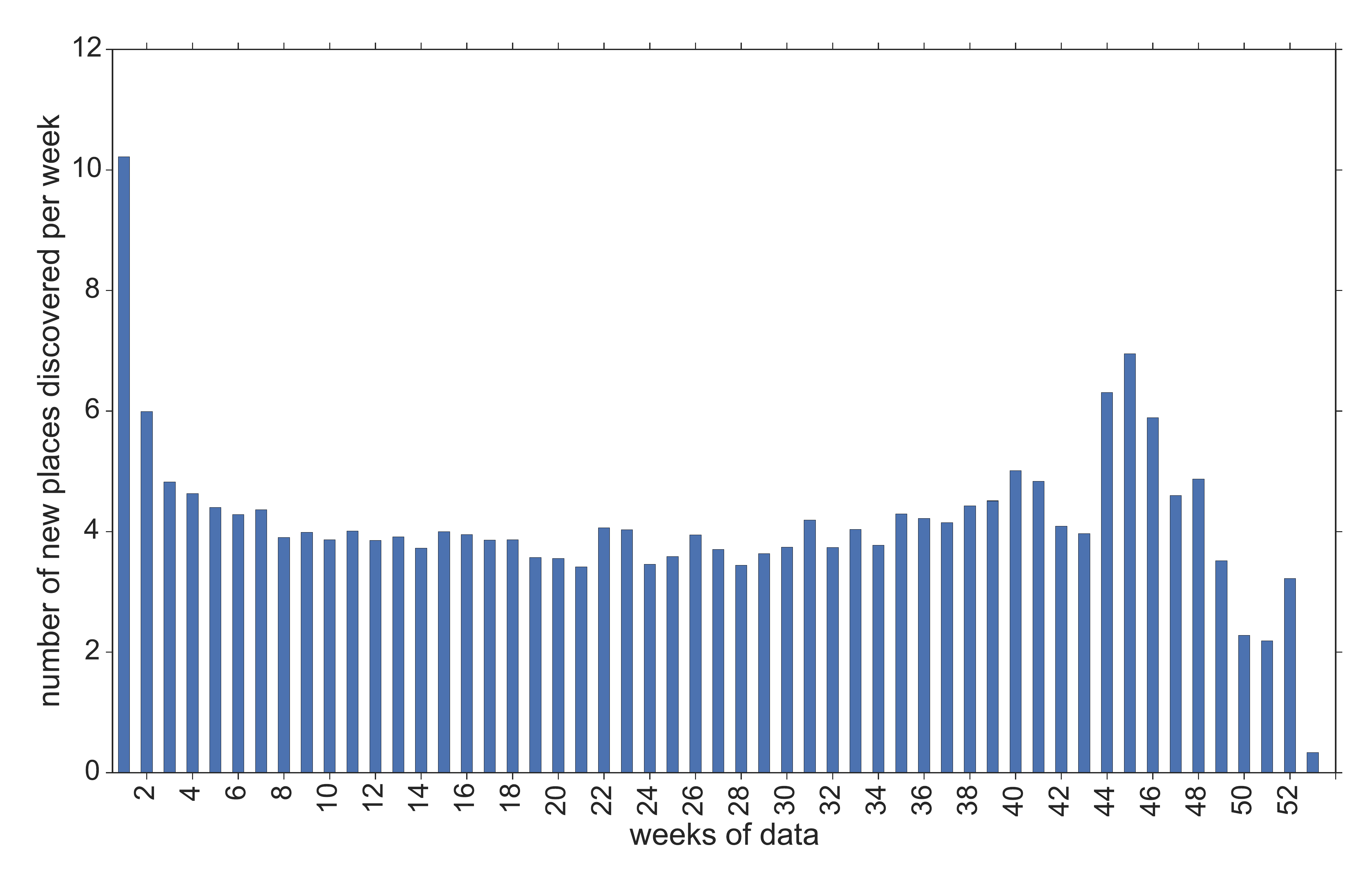}
\caption{Number of new visited (explored) places for each week, average by user. 
Surprisingly, the number of explored places does not decrease over time, but remains around 4.
This highlights the highly exploring behavior of our population.}
\label{fig:exploration_week}
\end{figure}

\begin{figure}[!h]
\centering
\includegraphics[width=1\columnwidth]{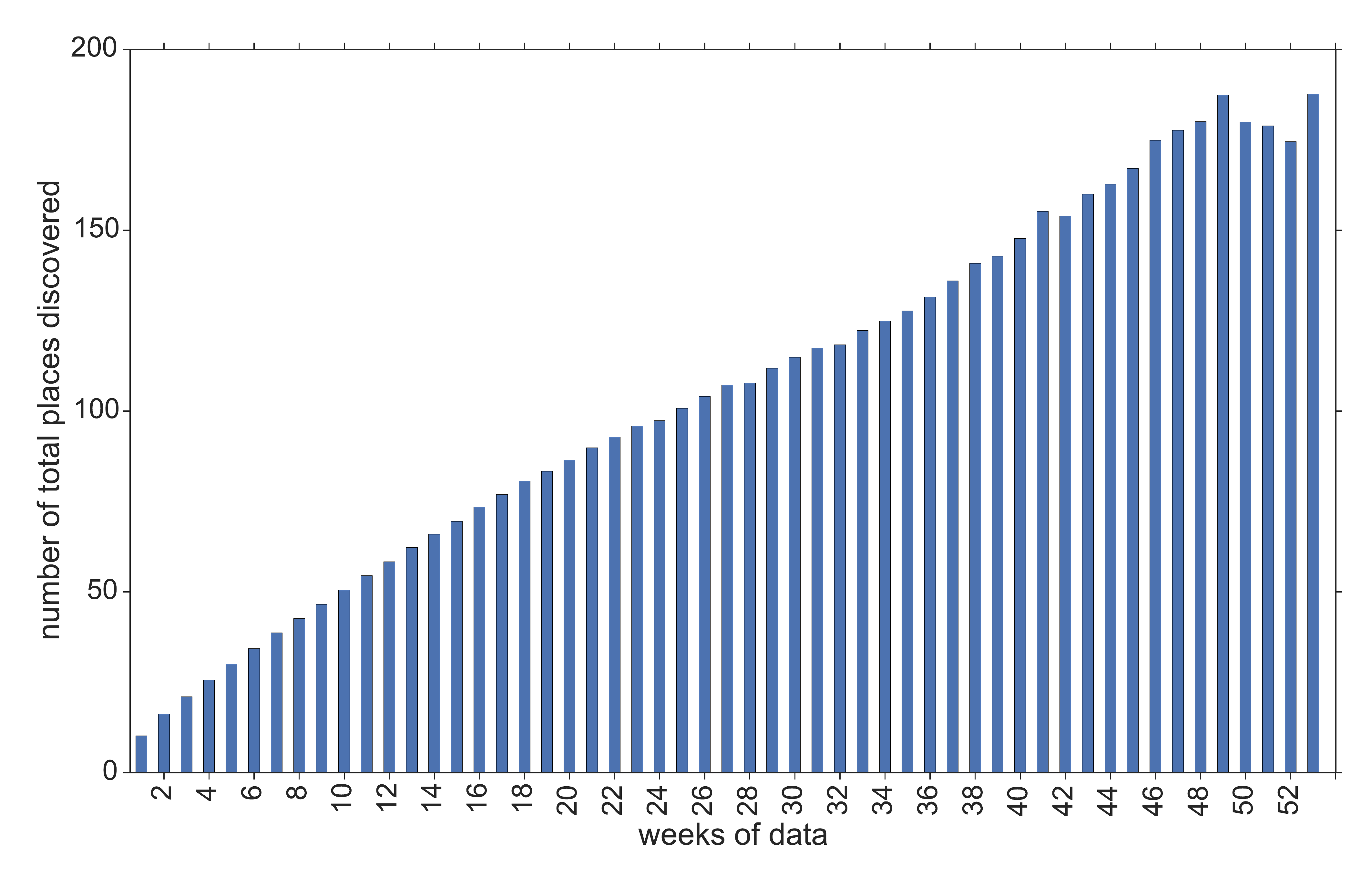}
\caption{Cumulated number of new visited (explored) places for each week, average by user. 
As consequence of the large amount of exploration, the number of possible places to visit increases steadily over time, reaching on average almost 200 in one year.
}
\label{fig:exploration_cumulated}
\end{figure}

The fact that a large fraction of stop-locations have never been seen before poses a challenge for the prediction task, since by construction any model that tries to predict a place from an alphabet of previously visited places will be unable to predict new, unseen symbols.
Moreover, another consequence of this frequent exploration is that the pool of possible places constantly grows over time and, given the longitudinal nature of our dataset, ends up being very large.
Fig.~\ref{fig:exploration_week} shows how the average number of new place explored per week remains approximately constant around 4, and consequently the total number of places keeps growing to hundreds of places (Fig.~\ref{fig:exploration_cumulated}).
This is a problem for the prediction task, as the number of possible places that the classifier needs to choose from increases constantly.

In fact if we measure the relation between the number of unique places per user and the performance of the best performing logistic regression model using Pearson's correlation coefficient, we find a quite strong negative correlation ($r = -0.478$, $p < 0.001$).
On the other hand we find no significant correlation for accuracy with period length or number of stops.

These facts suggest that the exploration phenomenon is a key reason for the relatively low accuracy of mobility prediction tasks at high spatial resolution.
Given the importance of exploration, we now consider a novel task in mobility prediction: exploration prediction.

%%%%%%%%%%%%%%%%%%%%%%%%%%%%%%%%%%
\subsection*{Exploration prediction}
%%%%%%%%%%%%%%%%%%%%%%%%%%%%%%%%%%

The exploration prediction task can be stated as follows: given a user's location history up to step $n$, will the stop at step $n+1$ be an exploration or a return?

The first question is: what should be the baseline model for the exploration prediction task?
Surprisingly, most literature on human mobility prediction has focused on next location prediction but has overlooked the exploration prediction problem, and to the best of our knowledge no suitable solution has been proposed for this task.
We therefore suggest, as a reasonable baseline, random guessing with probability equal to our prior knowledge of the fraction of explorations: $P(exploration) \approx 0.2$.

For our main model we use as before the logistic regression model with the same features constructed for the next place prediction model.
We also add two additional features: \textit{explore\_now} and \textit{explore\_before}, which capture if the current stop or the previous stops were explorations, respectively. 
The intuition for these is that multiple explorations may occur in a row, and therefore the current exploration may increase the likelihood for an exploration at the next stop.
As before, we fit each individual separately, and we perform an online prediction, that fits based on the data up to step $n$, and predicts exploration at step $n+1$.
We fit one logistic regression model for each of the single features, and a more complex model with all the features at once.

\begin{figure}[!h]
\centering
\includegraphics[width=1\columnwidth]{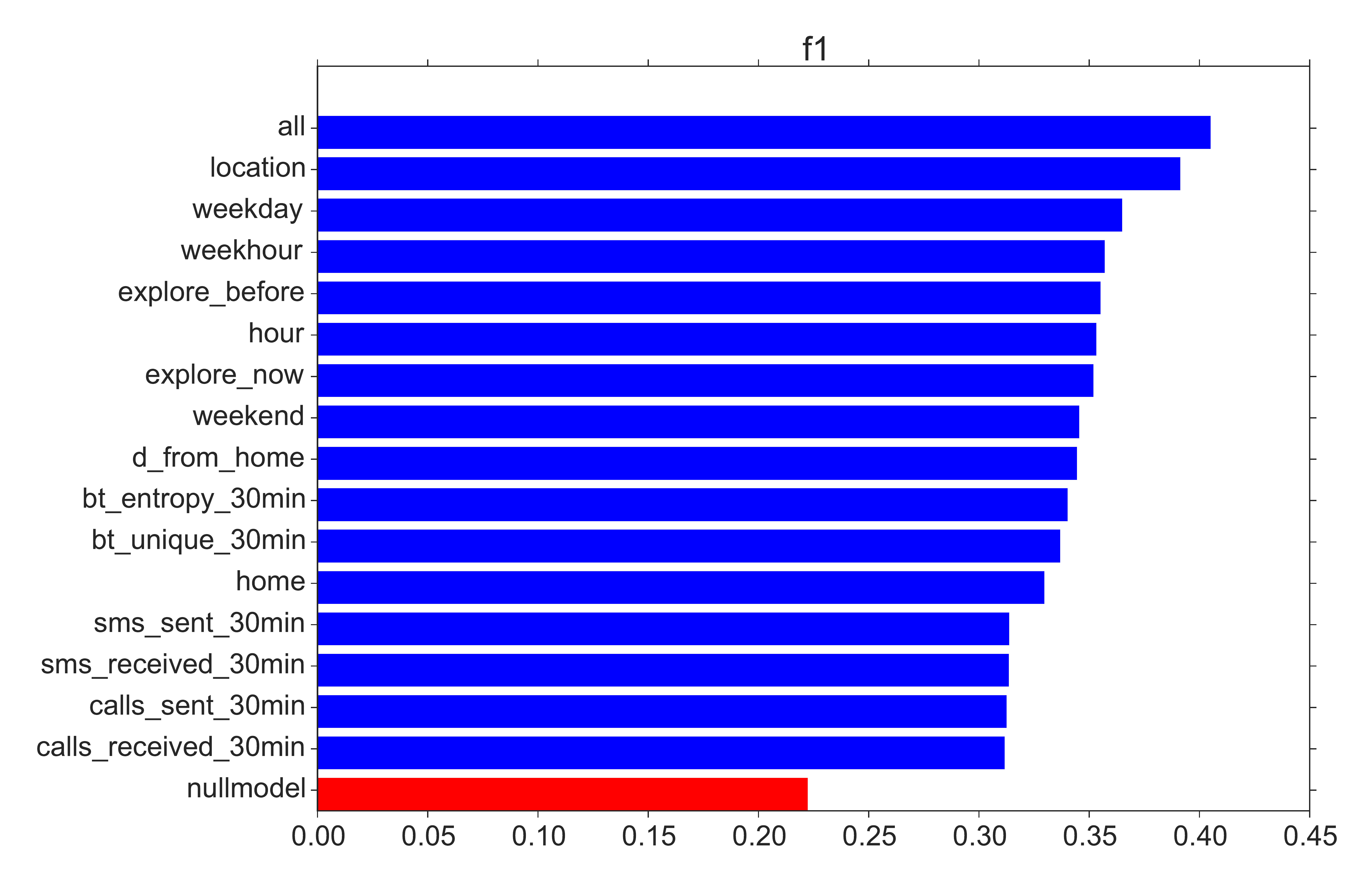}
\caption{Exploration prediction: $f_1$ score for all models.}
\label{fig:exploration-f1}
\end{figure}

Measuring the performance of these models requires a few considerations.
In this case, the classification problem is imbalanced, that is the number of positive cases (exploration) is much smaller than negative cases (return), as shown in Fig.~\ref{fig:p_exploration}.
This implies that accuracy is therefore not a good metric, since a classifier predicting always return (the most probable class) would have good performance, but would not be useful.
Instead we employ the $f_1$ score, which is the harmonic mean of precision (the fraction of correctly predicted explorations over all predicted explorations) and recall (the fraction of correctly predicted explorations over all true explorations).
Fig.~\ref{fig:exploration-f1} shows the results of the exploration prediction.

As we would expect, the model with the most complete set of features outperforms the others.
Among the single feature models, perhaps not surprisingly, the current location feature has the best performance. 
This finding can be explained by the role of some places as ``gateways'' for exploration such as public transport hubs (e.g.~central train station).
The individual features that perform also well are the time-related ones, in agreement with the intuition that exploration tends to happen to at specific times of the day or week.
The \textit{explore\_now} and \textit{explore\_before} also perform well, suggesting an element of burstiness in the exploratory behavior.
If we consider our best performing model, we find that it has average precision of 0.3 and recall of 0.65.
Overall the performance of this model is far from perfect, showing that the exploration prediction problem is a challenging one.

%%%%%%%%%%%%%%%%%%%%%%%%%%%%%%%%%%%%%%%%%%%%%%%%%%%%%%
\section*{Discussion}
%%%%%%%%%%%%%%%%%%%%%%%%%%%%%%%%%%%%%%%%%%%%%%%%%%%%%%

In this paper we first show that when interpreting results of predictive performance there are a number of factors that must be taken into consideration.
The problem formulation is the central factor what should be taken into account when interpreting predictability results, since e.g.~predicting the next time-bin is a very different (and much easier) task than predicting the next transition.
We show that the most challenging problem is the next-place prediction, which is arguably the most useful task for practical applications such as travel recommendations.
Another issue to be taken into account is the spatial resolution of the prediction, here we show how more coarse spatial precision results in an easier task.
Similarly the time resolution also has an effect on the predictive power.
We suggest that the factors described in this paper should be taken in consideration as context when comparing results from prediction models.

Other than the factors discussed above, we believe that one further reason for performance differences could be the demographics of the dataset.
The population under study is here composed by students that have no single workplace but tend to change multiple classes per week, even multiple times per day.
Moreover a younger population may have a more irregular schedule and more exploratory behavior.
Certainly more work is needed to conclusively link demographics and predictability.
For future directions, we suggest considering demographic factors when trying to characterize human mobility, as it has been done, for example, by linking changes in mobility patterns with unemployment status~\cite{toole2015tracking}.

We also discussed the issue of exploration, and we show how frequently new places are discovered.
Based on that, we show that the mechanism of exploration is an important part of human mobility and plays a role in next-place prediction.
Because any model that tries to predict a next place from a set of visited place will fail when an exploration occurs.
This problem has rarely been addressed in mobility prediction literature, which almost always assumes that the next place can be determined from the past history. 
Providing a full solution for next explored place prediction is beyond the scope of this work, and here we simply aim to stress the fact that the prediction of explorations is very different from the predictions to returns to known places.
Some previous work on next-place prediction using social information~\cite{sadilek2012finding, cho2011friendship} or nearby Points Of Interest~\cite{yumodeling} may be the starting point for investigating this problem. 

In this sense, we raise the question on whether the simple location history is enough for accurate next-place prediction.
As we have discussed, there are indeed a lot of regularities both in the sequence of visits, and in the daily and weekly temporal patterns of visitation.
However there are a lot of ``exceptions to the rules'', where schedules change, plans are canceled, and people run late.
We speculate that other channels such as email, social media, calendar, class schedule may be needed for achieving a satisfying accuracy in the prediction task.

%%%%%%%%%%%%%%%%%%%%%%%%%%%%%%%%%%%%%%%%%%%%%%%%%%%%%%
\section*{Acknowledgments}
%%%%%%%%%%%%%%%%%%%%%%%%%%%%%%%%%%%%%%%%%%%%%%%%%%%%%%

This work is funded in part by the High Resolution Networks project (The Villum Foundation), as well as Social Fabric (University of Copenhagen).


\begin{thebibliography}{10}

\bibitem{lane2011bewell}
Lane ND, Mohammod M, Lin M, Yang X, Lu H, Ali S, et~al.
\newblock BeWell: A Smartphone Application to Monitor, Model and Promote
  Wellbeing.
\newblock In: 5th International ICST Conference on Pervasive Computing
  Technologies for Healthcare; 2011. p. 23--26.

\bibitem{quercia2010recommending}
Quercia D, Lathia N, Calabrese F, Di~Lorenzo G, Crowcroft J.
\newblock Recommending Social Events from Mobile Phone Location Data.
\newblock In: Data Mining (ICDM), 2010 IEEE 10th International Conference on.
  IEEE; 2010. p. 971--976.

\bibitem{aalto2004bluetooth}
Aalto L, G{\"o}thlin N, Korhonen J, Ojala T.
\newblock Bluetooth And WAP Push Based Location-aware Mobile Advertising
  System.
\newblock In: Proceedings of the 2nd International Conference on Mobile
  systems, Applications, and Services. ACM; 2004. p. 49--58.

\bibitem{lu2012predictability}
Lu X, Bengtsson L, Holme P.
\newblock Predictability of Population Displacement after the 2010 Haiti
  Earthquake.
\newblock Proceedings of the National Academy of Sciences.
  2012;109(29):11576--11581.

\bibitem{ccolak2016understanding}
{\c{C}}olak S, Lima A, Gonz{\'a}lez MC.
\newblock Understanding Congested Travel in Urban Areas.
\newblock Nature Communications. 2016;7.

\bibitem{eagle2006reality}
Eagle N, Pentland A.
\newblock Reality Mining: Sensing Complex Social Systems.
\newblock Personal and Ubiquitous Computing. 2006;10(4):255--268.

\bibitem{liao2007learning}
Liao L, Patterson DJ, Fox D, Kautz H.
\newblock Learning and Inferring Transportation Routines.
\newblock Artificial Intelligence. 2007;171(5):311--331.

\bibitem{zheng2011computing}
Zheng Y, Zhou X.
\newblock Computing with Spatial Trajectories.
\newblock Springer Science \& Business Media; 2011.

\bibitem{arentze2000albatross}
Arentze T, Timmermans H.
\newblock Albatross: A Learning Based Transportation Oriented Simulation
  System.
\newblock Eirass Eindhoven; 2000.

\bibitem{balmer2008agent}
Balmer M, Meister K, Rieser M, Nagel K, Axhausen KW, Axhausen KW, et~al.
\newblock Agent-based Simulation of Travel Demand: Structure and Computational
  Performance of MATSim-t.
\newblock ETH, Eidgen{\"o}ssische Technische Hochschule Z{\"u}rich, IVT
  Institut f{\"u}r Verkehrsplanung und Transportsysteme; 2008.

\bibitem{goodchild2007citizens}
Goodchild MF.
\newblock Citizens as Sensors: The World of Volunteered Geography.
\newblock GeoJournal. 2007;69(4):211--221.

\bibitem{batty2013new}
Batty M.
\newblock The New Science of Cities.
\newblock MIT Press; 2013.

\bibitem{song2010limits}
Song C, Qu Z, Blumm N, Barab{\'a}si AL.
\newblock Limits of Predictability in Human Mobility.
\newblock Science. 2010;327(5968):1018--1021.

\bibitem{simini2012universal}
Simini F, Gonz{\'a}lez MC, Maritan A, Barab{\'a}si AL.
\newblock A Universal Model for Mobility and Migration Patterns.
\newblock Nature. 2012;484(7392):96--100.

\bibitem{schneider2013daily}
Schneider CM, Rudloff C, Bauer D, Gonz{\'a}lez MC.
\newblock Daily Travel Behavior: Lessons from a Week-long Survey for the
  Extraction of Human Mobility Motifs Related Information.
\newblock In: Proceedings of the 2nd ACM SIGKDD International Workshop on Urban
  Computing. ACM; 2013. p.~3.

\bibitem{lin2012predictability}
Lin M, Hsu WJ, Lee ZQ.
\newblock Predictability of Individuals' Mobility with High-resolution
  Positioning Data.
\newblock In: Proceedings of the 2012 ACM Conference on Ubiquitous Computing.
  ACM; 2012. p. 381--390.

\bibitem{smith2014refined}
Smith G, Wieser R, Goulding J, Barrack D.
\newblock A Refined Limit on the Predictability of Human Mobility.
\newblock In: Pervasive Computing and Communications (PerCom), 2014 IEEE
  International Conference on. IEEE; 2014. p. 88--94.

\bibitem{lu2013approaching}
Lu X, Wetter E, Bharti N, Tatem AJ, Bengtsson L.
\newblock Approaching the Limit of Predictability in Human Mobility.
\newblock Scientific Reports. 2013;3.

\bibitem{song2006evaluating}
Song L, Kotz D, Jain R, He X.
\newblock Evaluating Next-cell Predictors with Extensive Wi-fi Mobility Data.
\newblock Mobile Computing, IEEE Transactions on. 2006;5(12):1633--1649.

\bibitem{bapierre2011variable}
Bapierre H, Groh G, Theiner S.
\newblock A Variable Order Markov Model Approach for Mobility Prediction.
\newblock Pervasive Computing. 2011; p. 8--16.

\bibitem{zheng2010geolife}
Zheng Y, Xie X, Ma WY.
\newblock GeoLife: A Collaborative Social Networking Service among User,
  Location and Trajectory.
\newblock IEEE Data Eng Bull. 2010;33(2):32--39.

\bibitem{gao2012mobile}
Gao H, Tang J, Liu H.
\newblock Mobile Location Prediction in Spatio-temporal Context.
\newblock In: Nokia Mobile Data Challenge Workshop. vol.~41; 2012. p.~44.

\bibitem{laurila2012mobile}
Laurila JK, Gatica-Perez D, Aad I, Bornet O, Do TMT, Dousse O, et~al.
\newblock The Mobile Data Challenge: Big Data for Mobile Computing Research.
\newblock In: Pervasive Computing. EPFL-CONF-192489; 2012.

\bibitem{do2012contextual}
Do TMT, Gatica-Perez D.
\newblock Contextual Conditional Models for Smartphone-based Human Mobility
  Prediction.
\newblock In: Proceedings of the 2012 ACM Conference on Ubiquitous Computing.
  ACM; 2012. p. 163--172.

\bibitem{do2015probabilistic}
Do TMT, Dousse O, Miettinen M, Gatica-Perez D.
\newblock A Probabilistic Kernel Method for Human Mobility Prediction with
  Smartphones.
\newblock Pervasive and Mobile Computing. 2015;20:13--28.

\bibitem{scellato2011nextplace}
Scellato S, Musolesi M, Mascolo C, Latora V, Campbell AT.
\newblock Nextplace: A Spatio-temporal Prediction Framework for Pervasive
  Systems.
\newblock In: Pervasive Computing. Springer; 2011. p. 152--169.

\bibitem{sadilek2012far}
Sadilek A, Krumm J.
\newblock Far Out: Predicting Long-Term Human Mobility.
\newblock In: AAAI; 2012.

\bibitem{cho2011friendship}
Cho E, Myers SA, Leskovec J.
\newblock Friendship and Mobility: User Movement in Location-based Social
  Networks.
\newblock In: Proceedings of the 17th ACM SIGKDD International Conference on
  Knowledge Discovery and Data Mining. ACM; 2011. p. 1082--1090.

\bibitem{sadilek2012finding}
Sadilek A, Kautz H, Bigham JP.
\newblock Finding Your Friends and Following Them to Where You Are.
\newblock In: Proceedings of the fifth ACM International Conference on Web
  Search and Data Mining. ACM; 2012. p. 723--732.

\bibitem{stopczynski2014measuring}
Stopczynski A, Sekara V, Sapiezynski P, Cuttone A, Madsen MM, Larsen JE, et~al.
\newblock Measuring Large-scale Social Networks with High Resolution.
\newblock PloS one. 2014;9(4):e95978.

\bibitem{kang2004extracting}
Kang JH, Welbourne W, Stewart B, Borriello G.
\newblock Extracting Places from Traces of Locations.
\newblock In: Proceedings of the 2nd ACM International workshop on Wireless
  mobile applications and services on WLAN hotspots. ACM; 2004. p. 110--118.

\bibitem{zheng2010collaborative}
Zheng VW, Zheng Y, Xie X, Yang Q.
\newblock Collaborative Location and Activity Recommendations with GPS History
  Data.
\newblock In: Proceedings of the 19th International Conference on World Wide
  Web. ACM; 2010. p. 1029--1038.

\bibitem{thierry2013detecting}
Thierry B, Chaix B, Kestens Y.
\newblock Detecting Activity Locations from Raw Gps Data: A Novel Kernel-based
  Algorithm.
\newblock International Journal of Health Geographics. 2013;12(1):1.

\bibitem{zhou2007discovering}
Zhou C, Frankowski D, Ludford P, Shekhar S, Terveen L.
\newblock Discovering Personally Meaningful Places: An Interactive Clustering
  Approach.
\newblock ACM Transactions on Information Systems (TOIS). 2007;25(3):12.

\bibitem{zheng2009mining}
Zheng Y, Zhang L, Xie X, Ma WY.
\newblock Mining Interesting Locations and Travel Sequences from GPS
  Trajectories.
\newblock In: Proceedings of the 18th International Conference on World Wide
  Web. ACM; 2009. p. 791--800.

\bibitem{montoliu2010discovering}
Montoliu R, Gatica-Perez D.
\newblock Discovering Human Places of Interest from Multimodal Mobile Phone
  Data.
\newblock In: Proceedings of the 9th International Conference on Mobile and
  Ubiquitous Multimedia. ACM; 2010. p.~12.

\bibitem{ester1996density}
Ester M, Kriegel HP, Sander J, Xu X.
\newblock A Density-based Algorithm for Discovering Clusters in Large Spatial
  Databases with Noise.
\newblock AAAI Press; 1996. p. 226--231.

\bibitem{song2010modelling}
Song C, Koren T, Wang P, Barab{\'a}si AL.
\newblock Modelling the Scaling Properties of Human Mobility.
\newblock Nature Physics. 2010;6(10):818--823.

\bibitem{gonzalez2008understanding}
Gonzalez MC, Hidalgo CA, Barabasi AL.
\newblock Understanding Individual Human Mobility Patterns.
\newblock Nature. 2008;453(7196):779--782.

\bibitem{toole2015tracking}
Toole JL, Lin YR, Muehlegger E, Shoag D, Gonz{\'a}lez MC, Lazer D.
\newblock Tracking Employment Shocks Using Mobile Phone Data.
\newblock Journal of The Royal Society Interface. 2015;12(107):20150185.

\bibitem{yumodeling}
Yu C, Liu Y, Yao D, Yang LT, Jin H, Chen H, et~al.
\newblock Modeling User Activity Patterns for Next-Place Prediction.
\newblock IEEE Systems Journal. 2015;PP(99):1--12.
\newblock doi:{10.1109/JSYST.2015.2445919}.

\end{thebibliography}
\end{document}